\documentclass[twocolumn, trackchanges]{aastex63}

\usepackage{multirow}
\usepackage{asmath}
\usepackage{graphicx}
\usepackage{bm}
\usepackage[T1]{fontenc}

\newcommand{\myemail}{cornachi@usna.edu}
\newcommand{\Romannumeral}[1]{\uppercase\expandafter{\romannumeral #1\relax}}

\shorttitle{NIR and optical microlensing in Q0957 and SBS0909}
\shortauthors{Cornachione et al.}

\published{Nov 15, 2020}

\begin{document}

\title{Near infrared and optical continuum emission region size measurements in the gravitationally lensed quasars Q0957+561 and SBS0909+532}

\author[0000-0003-1012-4771]{Matthew A. Cornachione}
\affiliation{Department of Physics, United States Naval Academy, 572C Holloway Rd., Annapolis, MD 21402, USA}
\email{\myemail}

\author[0000-0003-2460-9999]{Christopher W. Morgan}
\affiliation{Department of Physics, United States Naval Academy, 572C Holloway Rd., Annapolis, MD 21402, USA}

\author{Hayden R. Burger}
\affiliation{Department of Physics, United States Naval Academy, 572C Holloway Rd., Annapolis, MD 21402, USA}

\author[0000-0003-3062-7835]{Vyacheslav N. Shalyapin}
\affiliation{O.Ya. Usikov Institute for Radiophysics and Electronics, National Academy of Sciences of Ukraine, 12 Acad. Proscury St., UA-61085 Kharkiv, Ukraine}
\affiliation{Departamento de F\'\i sica Moderna, Universidad de Cantabria, Avda. de Los Castros s/n, E-39005 Santander, Spain}
\affiliation{Institute of Astronomy of V.N. Karazin Kharkiv National University, Svobody Sq. 4, UA-61022 Kharkiv, Ukraine}

%\affiliation{O.Ya. Usikov Institute for Radiophysics and Electronics National Academy of Sciences of Ukraine 12 Acad. Proscury St., 61085 Kharkiv, Ukraine}
%\affiliation{Departamento de Física Moderna Universidad de Cantabria Avda. de Los Castros s/n, E-39005 Santander, Spain}

\author[/0000-0003-0110-834X]{Luis J. Goicoechea}
\affiliation{Departamento de F\'\i sica Moderna, Universidad de Cantabria, Avda. de Los Castros s/n, E-39005 Santander, Spain}

\author{Frederick J. Vrba}
\affiliation{United States Naval Observatory, Flagstaff Station, 10391 West Naval Observatory Road, Flagstaff, AZ 86005, USA}

\author{Scott E. Dahm}
\affiliation{United States Naval Observatory, Flagstaff Station, 10391 West Naval Observatory Road, Flagstaff, AZ 86005, USA}

\author{Trudy M. Tilleman}
\affiliation{United States Naval Observatory, Flagstaff Station, 10391 West Naval Observatory Road, Flagstaff, AZ 86005, USA}

\begin{abstract}
We present a microlensing analysis of updated light curves in three filters, $g$--band, $r$--band, and $H$--band, for the gravitationally lensed quasars Q0957+561 and SBS0909+532.  Both systems display prominent microlensing features which we analyze using our Bayesian Monte Carlo technique to constrain the quasar continuum emission region sizes in each band. We report sizes as half-light radii scaled to a $60\degr$ inclination angle. For Q0957+561 we measure $\log{(r_{1/2}/\text{cm})} = 16.54^{+0.33}_{-0.33}$, $16.66^{+0.37}_{-0.62}$, and 
$17.37^{+0.49}_{-0.40}$
%$18.01^{+0.86}_{-0.71}$
in $g$--, $r$--, and $H$--band respectively. For SBS0909+532 we measure $\log{(r_{1/2}/\text{cm})} = 15.83^{+0.33}_{-0.33}$, $16.21^{+0.37}_{-0.62}$, and 
$17.90^{+0.61}_{-0.63}$
%$17.93^{+0.59}_{-0.61}$
in $g$--, $r$--, and $H$--band respectively.  With size measurements in three bands spanning the quasar rest frame ultraviolet to optical, we can place constraints on the scaling of accretion disk size with wavelength, $r\propto\lambda^{1/\beta}$. In a joint analysis of both systems we find a slope shallower than that predicted by thin disk theory, $\beta = 0.35^{+0.16}_{-0.08}$, consistent with other constraints from multi-epoch microlensing studies.
% $\beta = 0.28^{+0.12}_{-0.07}$
\end{abstract}

\section{Introduction}

%[Key spin - multi-wavelength multi-epoch microlensing is a powerful tool to constrain quasar accretion disk size vs wavelength that has yet to be fully realized.  We present here the first H-band measurements in two systems and combined constraints on the slope with three-band measurements in two systems]

Quasars are difficult objects to study because their small physical sizes limit direct imaging and their immensely broad spectral energy distributions complicate theoretical modeling \citep{blae2004a, anto2013a}.
However, they play a valuable role in shaping our understanding of the universe. Their high luminosity enables observation at a very high redshift, giving us insight into the state of the early universe and galaxy formation \citep{boyl1998a, dima2005a}.  Furthermore, the strong-gravity environments at their centers provide a grounds for stringent testing of general relativity and alternate theories of gravitation \citep[e.g.][]{fabi2000a, char2017a, dai2019a}.  Quasars also have been used to measure the Hubble Constant \citep[e.g.][]{suyu2013a,bonv2017a,birr2019a, chen2019a, shaj2019a, wong2020a} and hold promise to be used as standard candles to constrain high-redshift cosmology \citep{wats2011a, luss2017a}. To fully unlock the potential of quasars, we must continue to refine our ability to measure their structure.

Multi-epoch quasar microlensing has emerged as one of the best means to measure the size of the quasar continuum emission region in gravitationally lensed quasars \citep{chan1979a, koch2004a}.  The flux in lensed quasar images is significantly magnified not only by the smooth mass profile of the lens galaxy, but also by individual lens galaxy stars.  This microlensing magnification varies, on time scales of years to decades, as the quasar moves relative to the lensing stars along our line of sight.  
%By monitoring lensed quasar image flux over long periods of time we can observe microlensing variability.  
From a careful analysis of multi-season light curves we can place constraints on the compact quasar continuum emission region size \citep{koch2004a, koch2006a, pool2007a, eige2008a, poin2008a, morg2010a, hain2012a, hain2013a, macl2015a, morg2018a, corn2020a}.  Microlensing measurements agree well with continuum emission region sizes measured through reverberation mapping \citep{pete2004a, bent2010a, mcha2014a, edel2015a, jian2016a, cack2018a, edel2019a}, but are anywhere from 2--4 times larger than sizes predicted from flux measurements when adopting the thin accretion disk model of \citet{shak1973a}.

In a complementary technique, single-epoch microlensing measures quasar sizes from simultaneous observations across multiple bands or from quasar spectra \citep{bate2008a, floy2009a, blac2011a, medi2011a, jime2012a, mott2012a, roja2014a, mott2017a, bate2018a, roja2020a}.  Although the single epoch method is sensitive to the unknown mean microlens mass, these sizes agree well with multi-epoch measurements and require significantly less observing time.
One additional strength of the single epoch method is that these multiple band observations provide estimates of microlensing-induced chromatic variability.  Chromatic variability allows constraints on the scaling of the quasar size with wavelength, $r\propto\lambda^{1/\beta}$.  The parameter $\beta$ characterizes the effective temperature profile slope of the underlying disk emission. Constraining $\beta$ may provide crucial insight into the physics governing quasar continuum emission and help distinguish between the many proposed accretion disk models \citep{corn2020b}.

Unfortunately not all chromatic microlensing measurements agree on the temperature profile slope.  While the standard \citet{shak1973a} thin disk model predicts $\beta=3/4$, some studies find a shallower slope of $\beta\approx0.5$ \citep{roja2014a, bate2018a} while others find a steeper slope of $\beta\approx 1.25$ \citep{mott2012a, jime2014a, roja2020a}.  Alternate means to measure the temperature profile slope through reverberation mapping \citep{cack2007a, jian2016a, cack2018a, edel2019a} do not yet tightly constrain the temperature profile slope, though their findings are generally consistent with the thin disk value.  Measurements from the quasar continuum spectral slope support a shallower value of $\beta \geq 0.57$ although the precise value depends on the host galaxy dust extinction \citep{davi2007a, bonn2013a, xie2016a}.

Multi-epoch chromatic microlensing measurements may be able to better determine the temperature profile slope since the underlying size measurements are more robust to uncertainty in the mean microlens mass.  However, because of the intensive observing requirements, most multi-epoch studies have focused on a single observing band.  Thus chromatic microlensing light curves have not been widely employed to estimate the temperature slope.  Several studies have analyzed multi-band light curves in the Einstein Cross (Q2237+0305) yielding estimates ranging from $\beta=0.75$ to $\beta=1.1$, consistent with the thin disk or mildly steeper \citep{eige2008a,muno2016a,goic2020a}.  \citet{poin2008a} analyzed HE1104-1805 in multiple bands, measuring $\beta=0.61^{+0.21}_{-0.17}$, also consistent with the thin disk.  Two band measurements in SBS 0909+532 and SDSS J0924+0219 yielded a shallower slope of $\beta\approx0.5$ but the error bars were too broad to exclude other models \citep{hain2013a, macl2015a}.  An alternate method of estimating $\beta$ from aggregated multi-epoch size measurements at $\lambda\approx 2500\,\text{\AA}$ strongly supported a shallower slope of $\beta < 0.56$, but these findings were model dependent \citep{corn2020b}.

We present here multi-epoch microlensing measurements three bands, $g$--, $r$--, and $H$--band, for two gravitationally lensed quasars Q0957+561 (hereafter Q0957) and SBS 0909+532 (hereafter SBS0909).  Q0957 has a particularly long history, starting with its discovery \citep{wals1979a} and early microlensing constraints published two decades ago \citep{refs2000a}.  Both systems are doubly imaged quasars with previous size measurements in the $r$--band with multi-epoch microlensing \citep{hain2012a,hain2013a}.   SBS0909 also has a $g$--band size measurement \citep{hain2013a}, though the uncertainties were broad.  Both quasars have also been analyzed in single epoch studies, enabling a ready comparison of the results \citep{medi2011a, mott2012a, jime2014a}.  We add the longest $H$--band light curves published for any lensed quasar, complemented by extremely long $g$-- and $r$--band light curves in both systems.  This provides a sample of widely-spaced wavelength measurements, allowing us to explore the power of the constraints that can be achieved with multi-epoch chromatic microlensing.

This paper is organized as follows.  We present the data from our monitoring campaign at the United States Naval Observatory (USNO) and Liverpool Robotic Telescope (LRT) in Section~\ref{sec:data}.
%, including a description of our infrared data reduction.
We describe our modeling, including strong lensing, time delays, and microlensing, in Section~\ref{sec:model}.  In Section~\ref{sec:micro} we explain our microlensing analysis techniques and we present our size and temperature slope measurements in Section~\ref{sec:results}.  Finally we discuss the implications of these findings in Section~\ref{sec:discuss}.

Throughout we adopt a cosmology of $\Omega_M = 0.3$, $\Omega_{\Lambda} = 0.7,$ and $H_0 = 70\,\text{km}\,\text{s}^{-1}\,\text{Mpc}^{-1}$ \citep{hins2009a}.

\section{Data}
\label{sec:data}

\begin{figure*}[t]
\centering
\includegraphics[width=1.0\textwidth]{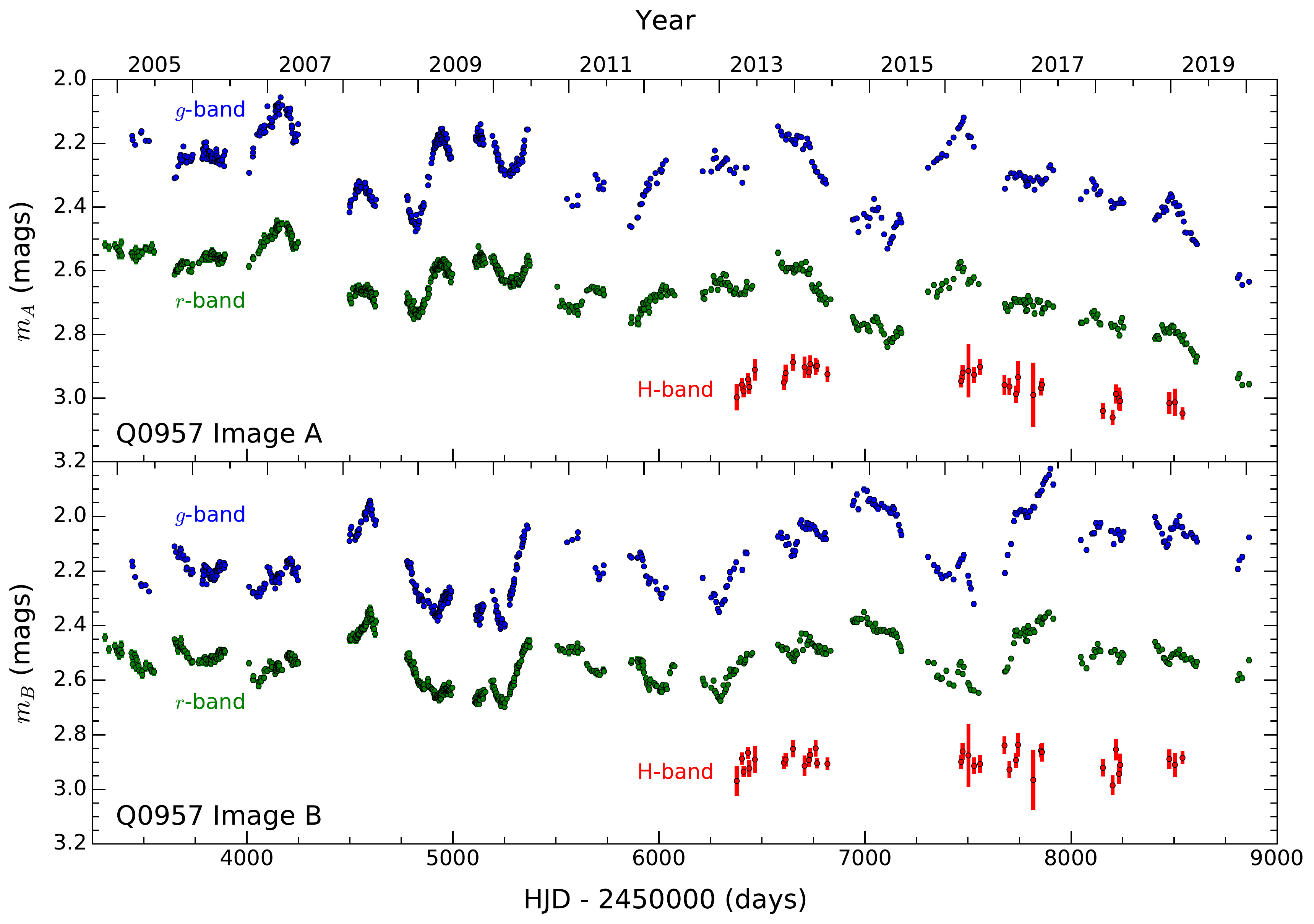}
\caption{Combined USNO and LRT light curves for Q0957.  Top: Image A light curves for $g$--band (blue), $r$--band (green), and $H$--band (red).  Bottom: Image B light curves for $g$--band (blue), $r$--band (green), and $H$--band (red).  The full $r$--band light curves are not shown and extend back to $\text{HJD} - 2450000 = 117$.  Light curves are vertically offset for clarity.}
\label{fig:lc_q0957}
\end{figure*}

\begin{figure*}[t]
\centering
\includegraphics[width=1.0\textwidth]{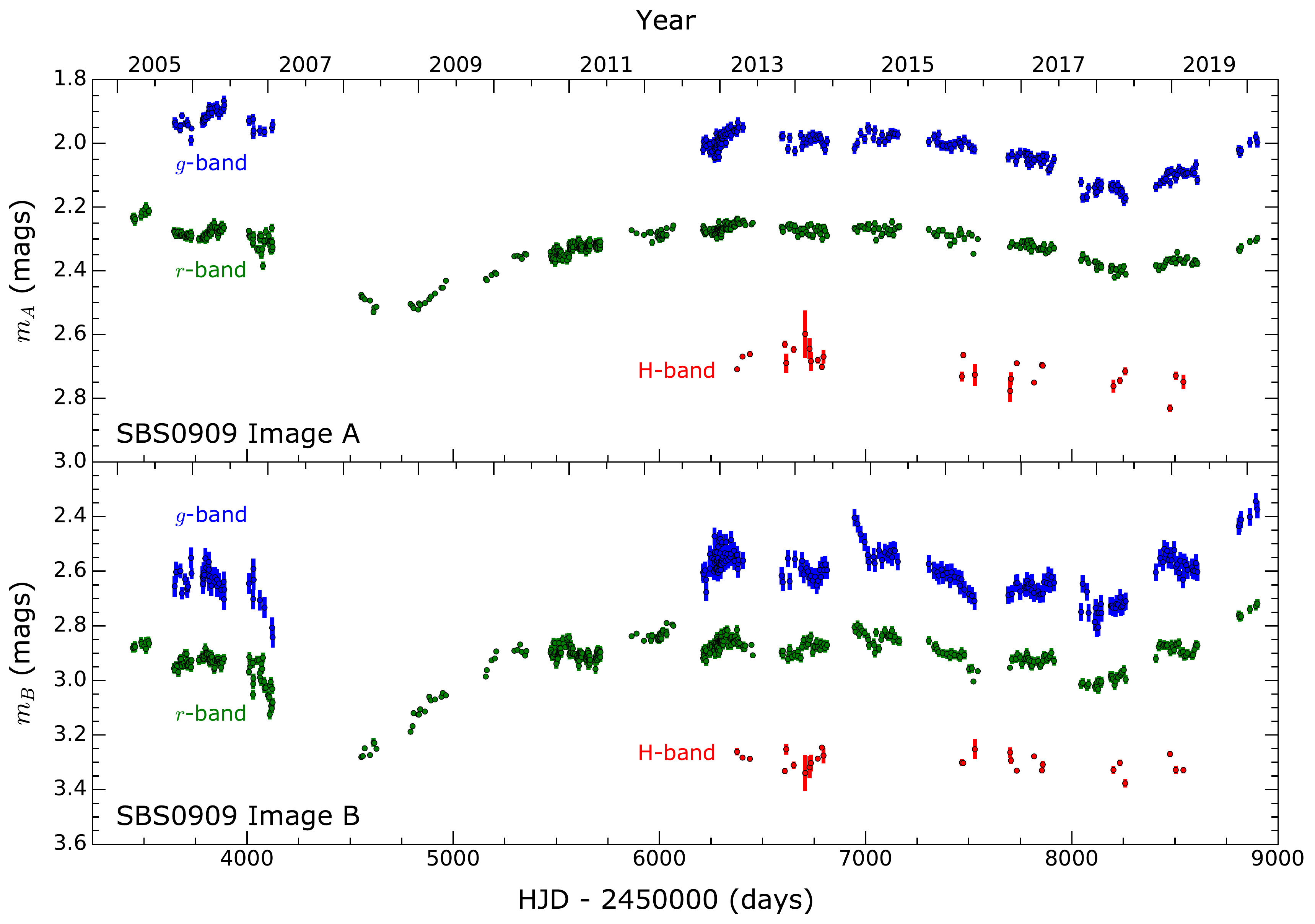} 
\caption{Combined USNO and LRT light curves for SBS0909.  Top: Image A light curves for $g$--band (blue), $r$--band (green), and $H$--band (red).  Bottom: Image B light curves for $g$--band (blue), $r$--band (green), and $H$--band (red). Light curves are vertically offset for clarity.}
\label{fig:lc_sbs0909}
\end{figure*}

\subsection{Monitoring Campaigns}

Our $r$--band observations are extensions of the programs described in \citet{hain2012a} and \citet{shal2012a} (for Q0957) and \citet{hain2013a} (for SBS0909).  These combine observations at the USNO, using the $1.55\,\text{m}$ Kaj Strand Astrometric Reflector at Flagstaff Station, and the $2.0\,\text{m}$ Liverpool Robotic Telescope \citep[LRT;][]{stee2004a} with effective wavelengths of $4868\,\text{\AA}$ and $6165\,\text{\AA}$ in $g$-- and $r$--band respectively.
For SBS0909 we also include one epoch from the Hiltner $2.4\,\text{m}$ telescope at the MDM Observatory and two epochs from the Wisconsin-Indiana-Yale-NOAO (WIYN) $3.5\,\text{m}$ telescope as detailed in \citet{hain2013a}.

The USNO $r$--band observations of Q0957 from 2008 to 2011 were previously published in \citet{hain2012a}.  The LRT $r$--band curves from January 2005 through May 2016 have also been previously reported \citep{shal2008a, hain2012a, shal2012a,gilm2018a}.  In addtion, \citet{hain2012a} and \citet{gilm2018a} included photometry from the Instituto de Astrof\`{i}sica de Canarias' (IAC) Teide Observatory from 1996 to 2005.  We extend these curves with 48 new nights from the USNO spanning November 2011 through April 2017. % (see Table~\ref{tab:q0957rband_usno}).
%The typical cadence was XXX although some portions of the light curve are more sparsely sampled.
 These are collected with three $300\,\text{s}$ subexposures per night using either the Tek2K CCD or the EEV CCD with pixel scales of $0\farcs33$ and $0\farcs18$ respectively.
%with a median seeing of XXX.
We also present 76 new nights from the LRT using the IO:O camera, with a pixel scale of $0\farcs30$, from October 2016 to January 2020 collected with a typical cadence of one $300\,\text{s}$ exposure every week. %[DETAILS FROM VYACHESLAV?]
Combining all of these observations we have an exceptional $r$--band light curve with 1248 epochs spanning twenty--four years, the latter part of which is shown in Figure~\ref{fig:lc_q0957}.

For SBS0909 the $r$--band analysis of \citet{hain2013a} used the USNO observations from 2008 to 2012 and LRT observations from 2005 to 2007 \citep{goic2008a} and 2010 to 2012. \citet{gilm2018a} published all LRT SBS0909 light curves through May 2016. We add an additional 44 nights from the USNO, with three $300\,\text{s}$ subexposures per night, covering February 2012 through April 2017, again using either the Tek2K CCD or the EEV CCD. % (see Table~\ref{tab:sbs0909rband_usno}).  
%As with Q0957, the typical cadence was XXX with sparser sampling from 20XX to 2017 and a median seeing of XXX.
For the updated LRT observations, we present 81 nights from November 2016 through January 2020.  These used the IO:O camera with a typical cadence of one $150\,\text{s}$ exposure every week.  Altogether we have a combined $r$--band light curve with 510 epochs spanning sixteen years, shown in Figure~\ref{fig:lc_sbs0909}.

We include $g$--band observations of both targets collected from the LRT.  For Q0957 the LRT $g$--band light curve from January 2005 through June 2010 was previously published in \citet{shal2008a} and \citet{shal2012a}.  We present 198 new nights from December 2010 to January 2020.  Before May 2012 these were collected with a single $120\,\text{s}$ exposure using the RATCam while later nights used one $150\,\text{s}$ exposure with the IO:O camera.  These exposures were concurrent with the recent $r$--band observations, and collected at a similar cadence of one observation every seven days.  The total 555 epoch light curve is shown in Figure~\ref{fig:lc_q0957}. For SBS0909 the LRT $g$--band light curve between 2005 and 2007 was previously analyzed in \citet{hain2013a}.  In this work we add 195 nights of LRT $g$--band observations from October 2012 to February 2020.  All exposures were collected with the IO:O camera at a weekly cadence with a single exposure per night of $350\,\text{s}$ in the first 2012--2013 season and $300\,\text{s}$ thereafter.  Our updated SBS0909 $g$--band light curve spans 238 epochs, shown in Figure~\ref{fig:lc_sbs0909}.

Finally, we include $H$--band light curves for both Q0957 and SBS0909.  All $H$--band exposures were collected at the USNO with the ASTROCAM \citep{fisc2003a}, a $1024\times1024$ pixel InSb detector with a pixel scale of $0\farcs{}366$ and $10\farcs{}$ dithering between subexposures.  For Q0957 we present 35 nights, with three $400\,\text{s}$ subexposures per night, from March 2013 to June 2014 and March 2016 to Februrary 2019. For SBS0909 we observed 36 nights, with three $440\,\text{s}$ subexposures per night, across the same date ranges, March 2013 to June 2014 and March 2016 to Februrary 2019.  Both quasars were monitored at the cadence of roughly one observation per month and a median seeing of $1\farcs0$.  The $H$--band light curves are included in Figures~\ref{fig:lc_q0957} and~\ref{fig:lc_sbs0909}% and despite a significant sky background the typical $H$--band signal to noise ratio (SNR) is XXX for both quasars.

\subsection{Data Reduction}

\begin{figure*}[t]
\centering
\includegraphics[width=1.0\textwidth]{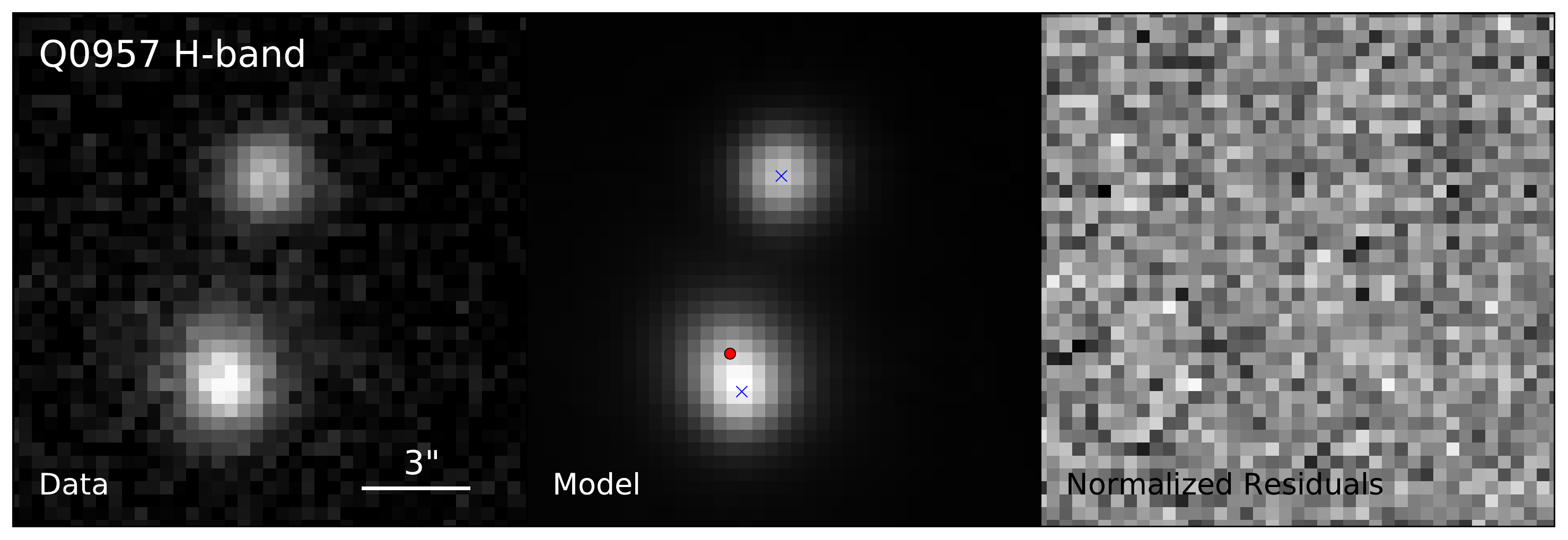} 
\includegraphics[width=1.0\textwidth]{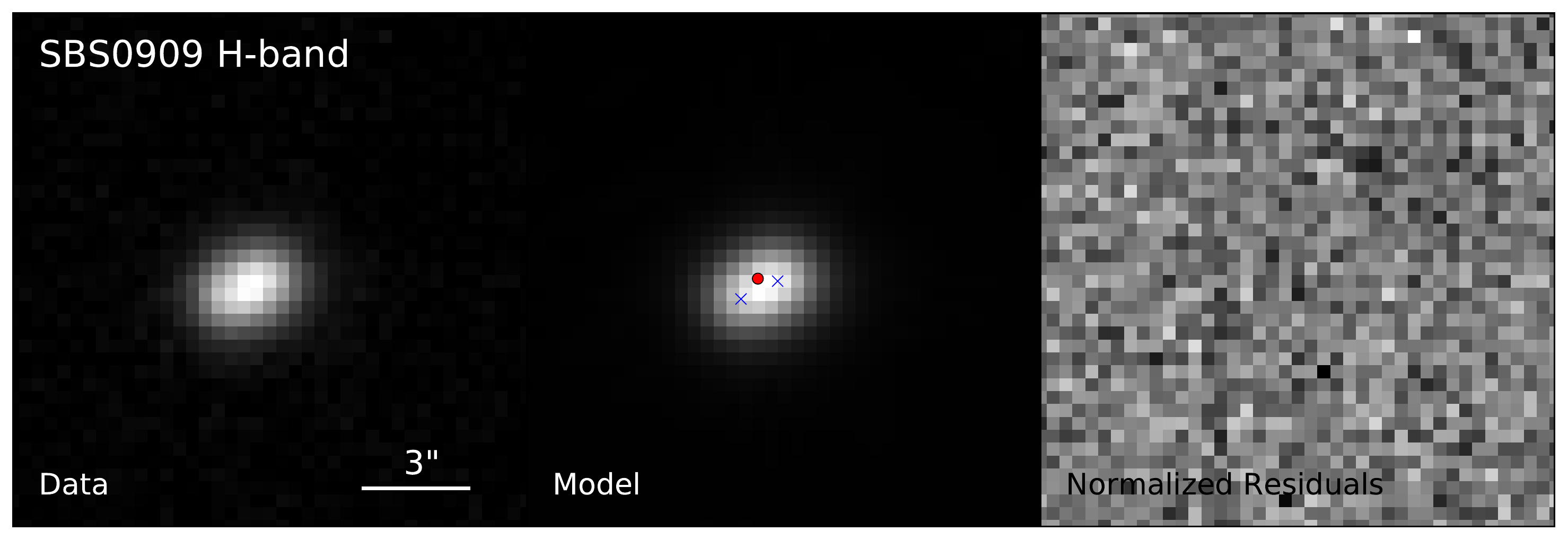} 
\caption{Data (with sky background subtracted), model, and error--normalized residuals for the $H$--band fits to Q0957 (top) and SBS0909 (bottom) of a typical subexposure from October 10, 2016.  In the center panel, the positions of the quasar image centroids are indicated with a blue $\times$ and the galaxy position with a red circle.}
\label{fig:img_fits}
\end{figure*}

\begin{deluxetable*}{cccccc}
\tablecaption{Q0957 combined $r$--band light curve from Teide Observatory, LRT, and USNO. \label{tab:q0957rband}}
\tablehead{
\colhead{HJD-2450000 (days)} & \colhead{Image A (mag)} & \colhead{Error A (mag)} & \colhead{Image B (mag)} & \colhead{Error B (mag)} & \colhead{Observatory}}
\startdata
117.0000 & 2.532 & 0.0230 & 2.480 & 0.022 & teide \\
118.0200 & 2.523 & 0.0230 & 2.472 & 0.022 & teide \\
119.0600 & 2.537 & 0.0230 & 2.487 & 0.022 & teide \\
120.0400 & 2.486 & 0.0220 & 2.432 & 0.021 & teide \\
122.0000 & 2.574 & 0.0230 & 2.514 & 0.022 & teide \\
\enddata
\tablecomments{This table is published in its entirety in machine-readable format. A portion is shown here for guidance regarding its form and content.}
\end{deluxetable*}

\begin{deluxetable*}{ccccc}
\tablecaption{Q0957 $g$--band light curve from the LRT. \label{tab:q0957gband}}
\tablehead{
\colhead{HJD-2450000 (days)} & \colhead{Image A (mag)} & \colhead{Error A (mag)} & \colhead{Image B (mag)} & \colhead{Error B (mag)}}
\startdata
3443.912 & 2.671 & 0.007 & 2.604 & 0.008 \\
3445.893 & 2.684 & 0.007 & 2.622 & 0.008 \\
3457.931 & 2.699 & 0.007 & 2.661 & 0.007 \\
3486.876 & 2.661 & 0.007 & 2.687 & 0.007 \\
3488.876 & 2.656 & 0.007 & 2.695 & 0.007 \\
\enddata
\tablecomments{This table is published in its entirety in machine-readable format. A portion is shown here for guidance regarding its form and content.}
\end{deluxetable*}

\begin{deluxetable*}{cccccc}
\tablecaption{SBS0909 combined $r$--band light curve from the MDM, WIYN, LRT and USNO. \label{tab:sbs0909rband}}
\tablehead{
\colhead{HJD-2450000 (days)} & \colhead{Image A (mag)} & \colhead{Error A (mag)} & \colhead{Image B (mag)} & \colhead{Error B (mag)} & \colhead{Observatory}}
\startdata
3038.664 & 2.303 & 0.009 & 2.899 & 0.013 & mdm \\
3048.620 & 2.283 & 0.007 & 2.910 & 0.007 & wiyn \\
3153.652 & 2.272 & 0.007 & 2.964 & 0.008 & wiyn \\
3445.888 & 2.233 & 0.014 & 2.881 & 0.018 & lrt \\
3452.928 & 2.232 & 0.014 & 2.880 & 0.018 & lrt \\
\enddata
\tablecomments{This table is published in its entirety in machine-readable format. A portion is shown here for guidance regarding its form and content.}
\end{deluxetable*}

\begin{deluxetable*}{ccccc}
\tablecaption{SBS0909 $g$--band light curve from the LRT. \label{tab:sbs0909gband}}
\tablehead{
\colhead{HJD-2450000 (days)} & \colhead{Image A (mag)} & \colhead{Error A (mag)} & \colhead{Image B (mag)} & \colhead{Error B (mag)}}
\startdata
3649.208 & 2.502 & 0.017 & 3.553 & 0.038 \\
3656.220 & 2.508 & 0.017 & 3.501 & 0.038 \\
3676.166 & 2.525 & 0.010 & 3.498 & 0.022 \\
3677.165 & 2.512 & 0.012 & 3.497 & 0.027 \\
3684.174 & 2.480 & 0.010 & 3.579 & 0.022 \\
\enddata
\tablecomments{This table is published in its entirety in machine-readable format. A portion is shown here for guidance regarding its form and content.}
\end{deluxetable*}

Details of the USNO $r$--band data reduction pipeline are described for Q0957 and SBS0909 in \citet{hain2012a} and \citet{hain2013a}.  In short, we stack the subexposures from each night to form a combined image.  We determine the point spread function (PSF) and flux normalization in each combined image using nearby reference stars (five stars for Q0957 and three stars for SBS0909).  Using the lens galaxy model and astrometry from \citet{leha2000a} we measure the individual image A and B flux at each epoch, holding the galaxy flux constant over all epochs.  For Q0957 we apply a small color offset of $\Delta m_{\text{Tek2K-EEV}} = 0.048\pm0.002\,\text{mag}$ between the two cameras, using $m_{\text{Tek2K}}$ as the reference magnitude, following the findings of \citet{hain2012a}.

The LRT reduction pipeline for Q0957 and SBS0909 is described in detail in \citet{gilm2018a}.  This is qualitatively similar to the USNO reduction though with a single nightly exposure no stacking was necessary.  Once again, we use a nearby reference star to determine the PSF for the exposure and the same star to calculate the flux normalization.  Using Hubble Space Telescope (\textsl{HST}) astrometry \citep{leha2000a}, we fit for an image centroid, background, and image fluxes with the IMFITFITS software \citep{mcle1998a,leha2000a}.  For Q0957 we included a contribution from the lens galaxy, held constant over all epochs, though for SBS0909 the lens galaxy was too faint in these bands to contribute significant flux.

We join the $r$--band USNO and LRT light curves using an offset of $14.43\,\text{mag}$ in Q0957 and $14.10\,\text{mag}$ in SBS0909, consistent with the offsets found in \citet{hain2012a, hain2013a}.  A sample of the $g$-- and $r$--band curves are given in Tables~\ref{tab:q0957rband}, \ref{tab:q0957gband}, \ref{tab:sbs0909rband}, and~\ref{tab:sbs0909gband} and these curves are provided in their entirety in electronic format. We also bin the light curves which reduces the run time of the microlensing analysis code.  We choose a binning of $20\,\text{days}$ for the long Q0957 $r$--band curve and $7\,\text{days}$ for the SBS0909 $r$--band curve and both $g$--band curves.  The microlensing variability is on a much longer time scale than our binning windows so this has a negligible influence on the size measurements.

For our $H$--band light curves we modify our data reduction pipeline to better account for the high sky background, brighter lens galaxy, and increased incidence of detector defects. We process each subexposure, first modeling the sky background as a smoothly varying shape with twelfth order Legendre polynomials along both pixel columns and rows, masking all stars, cosmic rays \citep[\texttt{astroscrappy\footnote{https://zenodo.org/record/1482019}},][]{vand2001b}, and detector defects.  We subtract this smooth term and then flat-field the image.

Instead of stacking the subexposures, we fit for the image A and B fluxes in each individual exposure, still holding the galaxy flux constant across all epochs.  We retain the USNO $r$--band reference stars for both Q0957 and SBS0909 and determine the PSF and flux normalization within each subexposure. We again use the astrometry from \citet{leha2000a} as an initial prediction of the relative quasar positions, but allow the image and galaxy centroids to vary within the measurement errors.  Example fits to Q0957 and SBS0909 are shown in Figure~\ref{fig:img_fits}.

For our final light curves, we calculate the flux each night as the weighted mean of the measured subexposure fluxes. To eliminate bad images, we reject any subexposures with a poor PSF fit to the reference stars ($\chi^2/N_{\text{dof}} > 1.5$).  We calculate the flux error for the night from the weighted mean of the flux errors in each subexposure $i$ using the standard formula $\sigma = \sqrt{\sum_i (1/\sigma_i^2)}$.  For some nights we found that the scatter in individual subexposure flux measurements significantly exceeds this error estimate.  In these nights, we include additional systematic uncertainty equal to the standard deviation of the subexposure fluxes.  This systematic contribution gives us a conservative error estimate in our final light curves.
The final curves are given in Tables~\ref{tab:q0957Hband_usno} and~\ref{tab:sbs0909Hband_usno}.

For Q0957 with an image separation of $6\farcs17$, the images are completely deblended, although we still apply a seeing cut of $1\farcs3$ because of blending between image B and the lens galaxy.  This eliminates 11 of the 103 total subexposures, although we retain at least one good subexposure for each night.  In SBS0909 the image A and B separation is only $1\farcs17$, requiring a more stringent seeing cut of $0\farcs9$ to exclude subexposures with significant flux sharing between images.  In total we cut 65 of the 103 subexposures, yielding 19 nights with at least one good subexposure.

\startlongtable
\begin{deluxetable*}{cccc}
\tablecaption{Q0957 $H$--band light curve from the USNO. \label{tab:q0957Hband_usno}}
\tablehead{
\colhead{HJD-2450000 (days)} & \colhead{Image A (mag)} & \colhead{Image B (mag)} & \colhead{Seeing (arcsec)}}
\startdata
6377.694 & $3.000\pm0.042$ & $3.073\pm0.058$ & $0.76$  \\
6402.747 & $2.961\pm0.021$ & $2.988\pm0.022$ & $0.98$  \\
6410.732 & $2.981\pm0.019$ & $3.046\pm0.020$ & $0.76$  \\
6432.705 & $2.947\pm0.020$ & $2.977\pm0.025$ & $1.19$  \\
6438.696 & $2.970\pm0.021$ & $3.018\pm0.024$ & $0.94$  \\
6465.656 & $2.914\pm0.034$ & $2.992\pm0.037$ & $1.14$  \\
6605.976 & $2.953\pm0.023$ & $3.010\pm0.025$ & $0.72$  \\
6615.012 & $2.922\pm0.026$ & $2.988\pm0.023$ & $0.97$  \\
6651.046 & $2.891\pm0.027$ & $2.959\pm0.024$ & $0.89$  \\
6706.856 & $2.904\pm0.034$ & $3.004\pm0.038$ & $0.88$  \\
6727.745 & $2.925\pm0.018$ & $2.993\pm0.022$ & $0.98$  \\
6734.712 & $2.893\pm0.029$ & $2.970\pm0.025$ & $1.13$  \\
6760.693 & $2.904\pm0.026$ & $2.953\pm0.030$ & $1.11$  \\
6767.738 & $2.898\pm0.019$ & $3.001\pm0.019$ & $0.79$  \\
6817.652 & $2.928\pm0.025$ & $3.015\pm0.025$ & $0.89$  \\
7466.747 & $2.952\pm0.019$ & $3.001\pm0.022$ & $0.82$  \\
7473.878 & $2.923\pm0.024$ & $2.966\pm0.029$ & $1.13$  \\
7501.765 & $2.919\pm0.083$ & $2.988\pm0.117$ & $1.26$  \\
7529.708 & $2.930\pm0.025$ & $3.015\pm0.030$ & $1.00$  \\
7558.653 & $2.908\pm0.025$ & $3.015\pm0.036$ & $1.00$  \\
7677.013 & $2.962\pm0.032$ & $2.950\pm0.033$ & $1.48$  \\
7701.018 & $2.966\pm0.027$ & $3.016\pm0.030$ & $1.14$  \\
7732.923 & $2.981\pm0.030$ & $2.982\pm0.024$ & $0.85$  \\
7742.909 & $2.938\pm0.051$ & $2.945\pm0.043$ & $1.47$  \\
7816.920 & $2.992\pm0.101$ & $3.042\pm0.106$ & $0.90$  \\
7853.792 & $2.973\pm0.024$ & $2.961\pm0.025$ & $1.07$  \\
7858.814 & $2.959\pm0.020$ & $2.958\pm0.026$ & $0.83$  \\
8154.874 & $3.042\pm0.026$ & $3.003\pm0.027$ & $0.83$  \\
8201.764 & $3.060\pm0.021$ & $3.047\pm0.032$ & $0.82$  \\
8217.764 & $2.991\pm0.030$ & $2.959\pm0.040$ & $1.32$  \\
8232.774 & $3.005\pm0.035$ & $3.055\pm0.037$ & $0.77$  \\
8238.743 & $3.013\pm0.030$ & $2.988\pm0.031$ & $1.14$  \\
8476.875 & $3.022\pm0.033$ & $2.993\pm0.027$ & $1.17$  \\
8503.994 & $3.017\pm0.043$ & $3.008\pm0.044$ & $0.88$  \\
8540.861 & $3.051\pm0.019$ & $2.994\pm0.024$ & $0.89$  \\
\hline
\enddata
\end{deluxetable*}

\startlongtable
\begin{deluxetable*}{cccc}
\tablecaption{SBS0909 $H$--band light curve from the USNO. \label{tab:sbs0909Hband_usno}}
\tablehead{
\colhead{HJD-2450000 (days)} & \colhead{Image A (mag)} & \colhead{Image B (mag)} & \colhead{Seeing  (arcsec)}}
\startdata
6352.728 & $2.254\pm0.222$ & $1.457\pm0.091$ & $1.58$  \\
6376.622 & $1.725\pm0.007$ & $1.854\pm0.030$ & $0.70$  \\
6402.701 & $1.710\pm0.004$ & $1.849\pm0.004$ & $0.86$  \\
6432.652 & $1.707\pm0.015$ & $1.846\pm0.013$ & $1.13$  \\
6438.648 & $1.707\pm0.008$ & $1.848\pm0.012$ & $0.90$  \\ 
6607.991 & $1.677\pm0.016$ & $1.905\pm0.031$ & $0.99$  \\
6614.967 & $1.692\pm0.014$ & $1.866\pm0.028$ & $0.98$  \\
6651.010 & $1.711\pm0.010$ & $1.866\pm0.010$ & $0.86$  \\
6706.803 & $1.748\pm0.030$ & $1.789\pm0.024$ & $0.98$  \\
6727.690 & $1.726\pm0.009$ & $1.832\pm0.008$ & $0.97$  \\
6734.655 & $1.711\pm0.023$ & $1.881\pm0.031$ & $0.93$  \\
6760.635 & $1.564\pm0.013$ & $2.059\pm0.022$ & $1.30$  \\
6767.696 & $1.714\pm0.009$ & $1.857\pm0.011$ & $0.80$  \\
6787.685 & $1.738\pm0.013$ & $1.815\pm0.011$ & $0.81$  \\
6795.637 & $1.709\pm0.020$ & $1.845\pm0.024$ & $0.85$  \\
6816.654 & $1.711\pm0.017$ & $1.847\pm0.015$ & $1.16$  \\
7466.699 & $1.749\pm0.006$ & $1.844\pm0.006$ & $0.79$  \\
7473.770 & $1.743\pm0.012$ & $1.843\pm0.007$ & $0.92$  \\
7501.719 & $1.748\pm0.014$ & $1.877\pm0.014$ & $1.07$  \\
7529.664 & $1.725\pm0.011$ & $1.845\pm0.012$ & $0.99$  \\
7676.970 & $1.672\pm0.018$ & $1.971\pm0.019$ & $1.31$  \\
7700.975 & $1.795\pm0.019$ & $1.832\pm0.023$ & $1.03$  \\
7704.920 & $1.762\pm0.019$ & $1.865\pm0.013$ & $0.79$  \\
7732.882 & $1.726\pm0.012$ & $1.906\pm0.013$ & $0.95$  \\
7742.862 & $2.532\pm0.210$ & $1.439\pm0.081$ & $1.44$  \\
7816.869 & $1.724\pm0.011$ & $1.894\pm0.011$ & $1.03$  \\
7854.723 & $1.746\pm0.010$ & $1.882\pm0.012$ & $0.90$  \\
7858.766 & $1.734\pm0.009$ & $1.876\pm0.010$ & $0.81$  \\
8201.689 & $1.808\pm0.018$ & $1.896\pm0.014$ & $0.88$  \\
8217.712 & $1.911\pm0.041$ & $1.798\pm0.036$ & $1.47$  \\
8232.735 & $1.780\pm0.010$ & $1.874\pm0.013$ & $0.86$  \\
8238.697 & $1.764\pm0.015$ & $1.906\pm0.018$ & $1.09$  \\
8258.665 & $1.777\pm0.015$ & $1.911\pm0.013$ & $0.92$  \\
8475.977 & $1.848\pm0.016$ & $1.851\pm0.016$ & $1.00$  \\
8503.947 & $1.796\pm0.021$ & $1.882\pm0.014$ & $0.91$  \\
8540.808 & $1.791\pm0.014$ & $1.885\pm0.010$ & $0.87$  \\
\hline
\enddata
\end{deluxetable*}

\section{Modeling}
\label{sec:model}

Microlensing analysis of gravitationally lensed quasars requires knowledge of the lens galaxy mass profile, the intrinsic quasar flux variability, and the magnification of stars in the lens galaxy.  Here we describe our approach to modeling each of these components. 

\subsection{Strong Lensing Models}
\label{sec:stronglens}

We adopt the same strong gravitational lensing models used in \citet{hain2012a} (Q0957) and \citet{hain2013a} (SBS0909).  In both systems we explore a sequence of models with a combination of external shear, a DeVaucouleurs baryonic matter profile, and a Navarro, Frenk, and White \citep[NFW][]{nava1997a} dark matter profile.  We parameterize this sequence with $f_{M/L}$ ranging from $0.1$ to $1.0$, where $f_{M/L}$ characterizes the magnitude of the DeVaucouleurs moment relative to a baryonic matter only model.  %These range from $0.1$ to $1.0$ in increments of $0.1$ where $f_{M/L}=1.0$ corresponds to a DeVaucouleur's only mass profile.
This model sequence allows us to analyze a range of stellar matter convergence to total convergence ratios, $\kappa_*/\kappa$ and shear, $\gamma$, values
%Our microlensing analysis marginalizes over this unknown convergence ratio
which \citet{sche2002a} show strongly affects the microlensing statistics.

For Q0957 our model sequence is based on the strong lensing models of \citet{fade2010a}.  In SBS0909, as explained in \citet{hain2013a}, strong lens modeling is complicated because the lens galaxy models of \citet{leha2000a} and \citet{slus2012a} are formally inconsistent.  \citet{hain2013a} explored microlensing statistics under both models and found that microlensing size measurements were not strongly affected by this uncertainty, although the \citet{slus2012a} models are harder to reconcile with the measured time delay.  We follow \citet{hain2013a} and adopt a blended model in our analysis, using the image and galaxy positions from \citet{leha2000a} and the lens galaxy shape from \citet{slus2012a}.  This blended model best matches the measured time delay and, in our model sequence, covers a wide range of possible stellar convergence values.

\subsection{Intrinsic Flux}
\label{sec:intrinsic}

Before running our analysis we must distinguish the intrinsic quasar variability from the extrinsic microlensing variability.
When the time delay between images A and B is known we can eliminate the common intrinsic variability by shifting the light curves to a common time basis and taking the ratio (or the magnitude difference).
Fortunately both systems have tightly constrained time delays from their long light curves, allowing us to readily find the difference light curves.

We adopt the time delay of $t_{AB} = 420\pm 2\,\text{days}$ (A leads B) from \citet{shal2012a} for Q0957 in the $r$--band.  This time delay is very precisely measured due to the %long light curve and 
high intrinsic variability during the observation period.  We confirm that the full, binned $g$--band light curve is consistent with this time delay using PyCS \citep{tewe2013a}.  We shift the curves by the time delay, using the leading image A as the reference curve, and linearly interpolate between nights.
%, excluding those more than 7 days into inter-season gaps.
For the Q0957 $H$--band measurements, given the large time delay and an observing gap between June 2014 and March 2016, we are left with only $5$ well-constrained nights in the difference light curve.  To maximize constraints for the microlensing analysis, we include extrapolated dates up to the system time delay on the end of the light curve and interpolated points in the observing gap.  For both the interpolated and extrapolated points, we adopt conservative error estimates based on the intrinsic quasar variability.  We model this variability using \texttt{celerite} \citep{fore2017a} to generate a damped random walk based on the observed power spectrum.

For SBS0909 we adopt the time delay of $t_{AB} = -50^{+2}_{-4} \,\text{days}$ (B leads A) from \citet{hain2013a}, formally consistent with the delay of $-49\pm 6\,\text{days}$ found in \citet{goic2008a}.  While \citet{eula2011a} found that the time delay of SBS0909 was weakly constrained, we used PyCS to analyze the longer $r$--band light curve and found a time delay of $-51 \pm 6\,\text{days}$, consistent with \citet{goic2008a} and \citet{hain2013a}.  As with Q0957, we shift the light curves by the time delay, using the leading image B as the reference curve, and linearly interpolate between nights.
%we are unable to get a strong constraint on the time delay. We explore the impact of time delays at $-30\,\text{days}$ and $-70\,\text{days}$ on the measured microlensing sizes (though the $-70\,\text{day}$ delay is very hard to reconcile with time delays predicted by strong lens models - see Figure~5 in \citet{hain2013a}).

\subsection{Microlensing Magnification Patterns}

Lens galaxy stars form a complex magnification pattern which we cannot observe directly but can model statistically based on the properties of the lens galaxy.  Our microlensing patterns are described in \citet{hain2012a} and \citet{hain2013a}.
These patterns use the values for stellar convergence $\kappa_*$, total convergence $\kappa$, and shear $\gamma$ produced by the lens model sequences described in Section~\ref{sec:stronglens}.
These sets include $40\times 10=400$ possible magnification patterns, formed from forty separate trials each using the range of 10 different $f_{M/L}$ values.
This samples a wide range of stellar lensing statistics.

Within each pattern, we include a range of stellar masses following $dN/dM \propto M^{-1.3}$ with a ratio of maximum to minimum mass of 50 following \citet{goul2000a}.  The mean stellar microlens mass $\langle M_*/M_{\odot} \rangle$ is left as a free parameter, determined during the microlensing analysis. While the Salpeter mass function is a more common choice, we follow \citet{poin2010a} in choosing the bulge mass function of \citet{goul2000a} which we expect to dominate near the galactic center region most relevant to strong lensing.  \citet{cong2007a} found that the microlensing magnification statistics are insensitive to the slope of the mass function and depend more on the dynamic mass range, where our range of 50 is comparable to the broadest range they examined.  Thus changes in the stellar mass function are unlikely to have a significant impact on the measured microlensing sizes.
These patterns are cast on a $8192\times 8192$ pixel grid.  We scale these to give a minimum size resolution at one pixel of $0.005 R_E$ and a maximum size resolution of $40 R_E$.  The Einstein size unit, $R_E = D_{OS}\theta_E$, is the product of the source (quasar) angular diameter distance and the Einstein radius and scales with the mean microlens mass as $\theta_E \propto \langle M_*/M_{\odot} \rangle^{1/2}$.   This wide dynamic range is important to obtain simultaneous size constraints in three different bands, where quasar continuum emission region sizes can vary by more than an order of magnitude.

  %These patterns provide possible realizations of the lens galaxy stellar magnifications projected onto the source plane.

%, although past microlensing studies have shown that the microlensing sizes are not strongly senstive to the shape of the mass function (citations maybe).  Indeed other microlensing studies choose to adopt a single mass (citations) without a loss of generality.  This mass function is characterized by one unknown, the median mass of the lens galaxy star, which we estimate during our microlensing analysis.

\section{Microlensing Analysis}
\label{sec:micro}

\begin{deluxetable*}{lcccc}
\tablecaption{Number of microlensing trial runs and good runs satisfying the given $\chi^2$ threshold for each analysis. \label{tab:nruns}}
\tablehead{
\colhead{Quasar} & \colhead{Analysis} & \colhead{$N_{\text{trials}}$} & \colhead{$\chi^2/N_{\text{dof}}$ threshold} & \colhead{$N_{\text{good}}$}}
\startdata
Q0957 & $g$--band only &  $4\times10^9$ & 2.1 & $1.7\times10^6$ \\
 &  $r$--band only & $4\times10^{10}$ & 2.1 & $1.3\times10^7$ \\
 &  $r/g$--band joint &  $2.7\times10^8$ & 1.9 & $1.2\times10^6$ \\
 &  $r$/$H$--band joint & $2.7\times10^8$ & 1.9 & $3.1\times10^7$ \\
 &  $g$/$H$--band joint & $3.6\times10^7$ & 2.1 & $5.0\times10^6$ \\
 \hline
SBS0909 & $g$--band only &  $4\times10^9$ & 2.4 & $2.6\times10^6$ \\
 &  $r$--band only & $4\times10^{10}$ & 1.9 & $2.5\times10^5$ \\
 &  $r/g$--band joint &  $5.2\times10^6$ & 3.0 & $827$ \\
 &  $r$/$H$--band joint & $5.2\times10^6$ & 3.0 & $5.1\times10^4$ \\
\enddata
\end{deluxetable*}

Our microlensing analysis follows the method detailed in \citet{koch2004a} and used in the previous studies of Q0957 \citep{hain2012a} and SBS0909 \citep{hain2013a}.  We run single band analyses on the $r$-- and $g$--band difference light curves.  For the $H$--band we run a joint size analysis with the long $r$--band curves. We also run a joint analysis combining the $r$-- and $g$--bands to check for consistency with the single band findings.

\subsection{Single Band Analysis}
\label{sec:ml_single}

%[Pass a range of sizes across the patterns at a range of velocities.  Impose $\chi^2$ cut]

%[Take the output and marginalize over the different variables to return a range of posteriors on values of interest (corner plot?)]

%In both Q0957 and SBS0909 our longest light curves are in the $r$-band.  We analyze these first.
%To model a finite quasar continuum emission region size, 
The quasar continuum flux magnification varies as the quasar moves across the screen of stellar lenses relative to our line of sight.  To analyze this variability, we first convolve each magnification pattern with a Gaussian kernel representing the quasar shape.  Our analysis uses a Gaussian convolution kernel for simplicity and speed of calculation because \citet{mort2005a} showed that the measured microlensing half-light sizes are weakly sensitive to the source shape, a conclusion supported in other studies \citep{koch2004a, cong2007a}.  We explore 21 different scale sizes of the Gaussian kernel, covering Einstein size units of $\log(\hat{r}_s) = [14.5,18.0]\langle M_*/M_{\odot} \rangle^{1/2}\,\text{cm}$ for Q0957 and $\log(\hat{r}_s) = [14.5,18.5]\langle M_*/M_{\odot} \rangle^{1/2}\,\text{cm}$ for SBS0909.  While the upper limit of $\log(\hat{r}_s)$ in SBS0909 slightly exceeds the size scale of the magnification patterns, we found this does not impact any of our key findings.  Our Monte Carlo analysis samples possible trajectories for both image A and B across these convolved patterns.  We randomize starting points, $(x_i,y_i)$, and directions, $\phi$, and choose effective source velocities, in Einstein units, from $\hat{v}_e = [10^1, 10^6]\langle M_*/M_{\odot} \rangle^{1/2}\,\text{km}\,\text{s}^{-1}$.  For the maximal $\hat{v}_e$, given our long light curves, the trajectories will cross an edge of the finite magnification pattern.  Our microlensing code includes periodic boundary conditions to allow for this.  We find, however, that both systems favor $\hat{v}_e < 2\times10^4\langle M_*/M_{\odot} \rangle^{1/2}\,\text{km}\,\text{s}^{-1}$, so edge effects do not impact our reported measurements of $r_{1/2}$.  
%in Q0957 and $\hat{r}_s = [14.5,18.0]$ in increments of XXX in SBS0909
We then compare these microlensing curves to the time-delay shifted difference curves described in Section~\ref{sec:intrinsic}, allowing for source optimization as in \citet{koch2004a}, and evaluate the quality of fit using the $\chi^2$ metric.  The degrees of freedom, $N_{\text{dof}}$, are given by the number of points in the difference curve minus the number of microlensing variables ($x_i,y_i,\phi,\hat{v}_e,\hat{r}_s,\kappa_*/\kappa$).
For Q0957 we add an overall flux ratio uncertainty of $0.1\,\text{mags}$ between images A and B.  This effectively accounts for errors in the strong lens models, any lens galaxy substructure magnification, differential dust extinction, and microlensing.  Because the strong lens models of SBS0909 are less certain (see Section~\ref{sec:stronglens}), we adopt a flux ratio uncertainty of $0.2\,\text{mags}$ in this system.

To speed up computational time and reduce output file sizes we impose a $\chi^2$ threshold in each system.  We adopted $r$--band constraints of $\chi^2/\text{N}_{\text{dof}} \leq 2.1$ for Q0957 and $\chi^2/\text{N}_{\text{dof}} \leq 2.4$ for SBS0909 and $g$--band constraints of $\chi^2/\text{N}_{\text{dof}} \leq 2.1$ for Q0957 and $\chi^2/\text{N}_{\text{dof}} \leq 1.9$ for SBS0909.  We select these thresholds by examining the total probability at each $\chi^2/\text{N}_{\text{dof}}$, which is the product of the number of good fits at a given $\chi^2$ times the probability from \citet{koch2004a}
\begin{equation}
P(\chi^2) \propto \Gamma\left[ \frac{N_{\text{dof}}-2}{2},\frac{\chi^2}{2f_0^2}\right].
\end{equation}
where $\Gamma$ is the incomplete gamma function and $f$ is a magnitude error scaling factor.  We set $f_0=1$ as in \citet{koch2004a}.
Solutions with $\chi^2/\text{N}_{\text{dof}}$ above these thresholds contribute negligible statistical weight ($<{\sim} 0.1$\%) toward the final solution.

Upon conclusion of the runs, we convert from Einstein units to physical units following \citet{koch2004a} and \citet{mosq2011a}.  This method uses cosmological models to constrain the physical velocity, $v_e$, which permits a determination of the mean microlensing mass $\langle M_*/M_{\odot} \rangle$ from the distribution of effective source velocities, $\hat{v}_e$.  We marginalize over the microlens mass, $f_{M/L}$, and trajectory variables $(v_e, \phi, x_i, y_i)$ to convert the distribution of Einstein sizes, $\hat{r}_s$, found in our analysis to a posterior distribution on the physical size of the quasar continuum emission region, $r_s$.

%Our analysis uses a Gaussian convolution kernel for simplicity and speed of calculation because \citet{mort2005a} showed that the measured microlensing sizes are weakly sensitive to the shape of the object.  Instead the half light radius dominates, which, for the Gaussian kernel is .  After conclusion of the run, we find the half-light radius from the Gaussian scale size, $r_{1/2} = 1.18r_{s,G}$.  This can be re-scaled to a fiducial thin disk size using $r_{s,thin} = r_{1/2}/2.44$.

We perform the Monte Carlo runs across the 400 trial magnification patterns with the United States Naval Academy high performance cluster, TheARC \footnote{https://www.usna.edu/ARCS/index.php}.
The number of runs performed for each system are shown in Table~\ref{tab:nruns} and are sufficiently large such that variations in the number of runs have a negligible impact on the measured sizes.
%We performed four separate runs, analyzing the $r$--band and $g$--band curves for both Q0957 and SBS0909. For Q0957 we found $1.3\times10^7$ curves that satisfy the $\chi^2$ threshold for the $r$--band and $1.7\times10^5$ curves for the $g$--band.  For SBS0909 we found $2.5\times10^5$ curves that satisfy the $\chi^2$ threshold for the $r$--band and XXX curves for the $g$--band.
The SBS0909 $r$--band has significantly fewer runs within the $\chi^2$ threshold than Q0957 because of a very high magnitude microlensing event from 2007 to 2010.  Though this feature is harder to fit, it provides strong size constraints.

\subsection{Two Band Joint Analysis}
\label{sec:ml_joint}

The $H$--band light curves are too sparse to return a tight constraint when analyzed alone.  Following the procedure used for X--rays \citep[e.g.][]{morg2008a,dai2010a} and short $g$--band curves \citep{hain2013a, macl2015a}, we perform a joint optical and $H$--band analysis.  This means we only fit the $H$--band difference curves to the solutions that satisfy the $\chi^2$ threshold of the $r$--band runs (see Table~\ref{tab:nruns}).

For each good $r$--band solution, we repeat the analysis described in Section~\ref{sec:ml_single} at 21 steps in quasar size. We explore the same size range in $H$--band of $\log(\hat{r}_s) = [14.5,18.5]\langle M_*/M_{\odot} \rangle^{1/2}\,\text{cm}$ for SBS0909.  For Q0957 we are able to compress this range, retaining all valid solutions, to $\log(\hat{r}_s) = [15.1,17.75]\langle M_*/M_{\odot} \rangle^{1/2}\,\text{cm}$, giving a finer size resolution. In the joint analysis, we use the same trajectories $(v_e, \phi, x_i, y_i)$ from the good $r$--band solutions rather than exploring a random set.  We assign equal statistical weight to the difference curve fits from each band and use an overall threshold of $\chi^2/\text{N}_{\text{dof}} \leq 1.9$ in Q0957 and $\chi^2/\text{N}_{\text{dof}} \leq 3.0$ in SBS0909.  We repeat the Bayesian posterior analysis on the results, also marginalizing over the $r$--band sizes.

A portion of the $H$--band flux is likely composed of scattered emission or host galaxy stars \citep{vand2001a, glik2006a, hern2016a}.  This large-scale diluting flux is not significantly microlensed.  In previous microlensing studies, the measured source size has been found to decrease as flux dilution increases \citep[e.g.][]{dai2010a}.  We seek to quantify the impact of this dilution in our joint $H$--band analysis.

Based on the spectral catalog published by \citet{cald2017a}, we estimate that the host galaxy can contribute up to 50\% of the total flux around $6800\,\text{\AA}$.  However \citet{shen2016a} found that host galaxy contamination decreases significantly for high luminosity quasars. Since both Q0957 and SBS0909 have bolometric luminosities of $L_{\text{bol}} \approx 10^{46}\,\text{erg}\,\text{s}^{-1}$ \citep{asse2011a}, we expect a much smaller diluting fraction. Indeed we find that if we restrict to \citet{cald2017a} quasars with $L_{\text{bol}} > 10^{45}\,\text{erg}\,\text{s}^{-1}$ we find only 10\% host galaxy contamination at $6800\,\text{\AA}$.  This fraction is roughly consistent with the host galaxy contamination reported in recent near infrared catalogs \citep{glik2006a, hern2016a}.  The $6563\,\text{\AA}$ H$\alpha$ line can also contribute up to 25\% to the total flux in these systems that will not be strongly microlensed. Based on these considerations, we choose to investigate the cases of 20\% flux dilution as a typical fraction, and 50\% flux dilution as an extreme upper bound.

Although we analyzed the $g$--band light curves independently, we also run a joint $r$-- and $g$--band analysis.  This provides a check on the findings of the two independent runs and, in principle, the tightest constraints on both sizes.  We again explore the same range of $\log(\hat{r}_{s})$ and keep all solutions with $\chi^2/\text{N}_{\text{dof}} \leq 3.0$.
For Q0957, with a robust $g$--band size measurement, we also computed the sizes under a joint $g$-- and $H$--band analysis for comparison.

\section{Results}
\label{sec:results}

\begin{figure*}[t]
\centering
\includegraphics[width=0.9\textwidth]{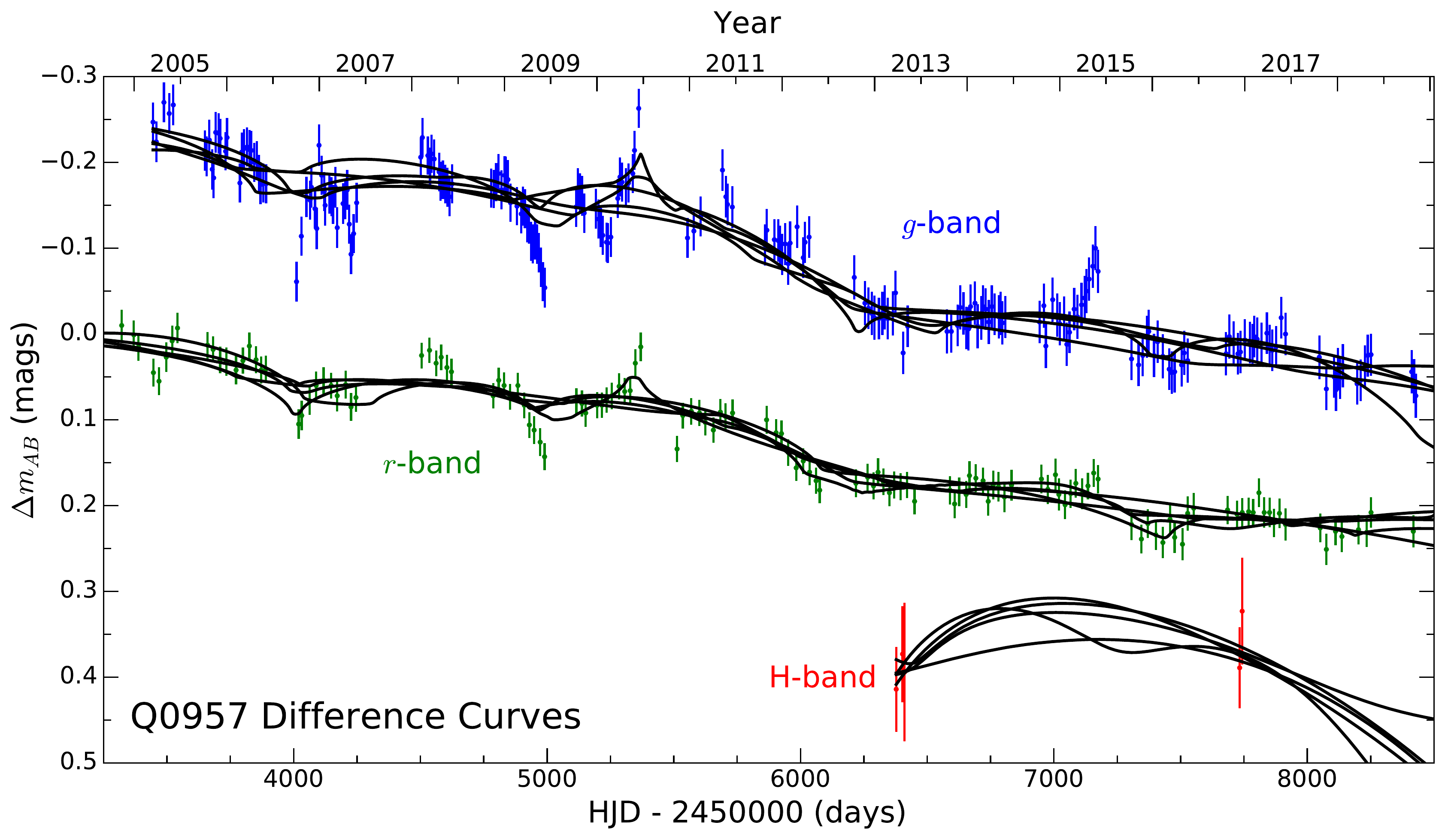} 
\includegraphics[width=0.9\textwidth]{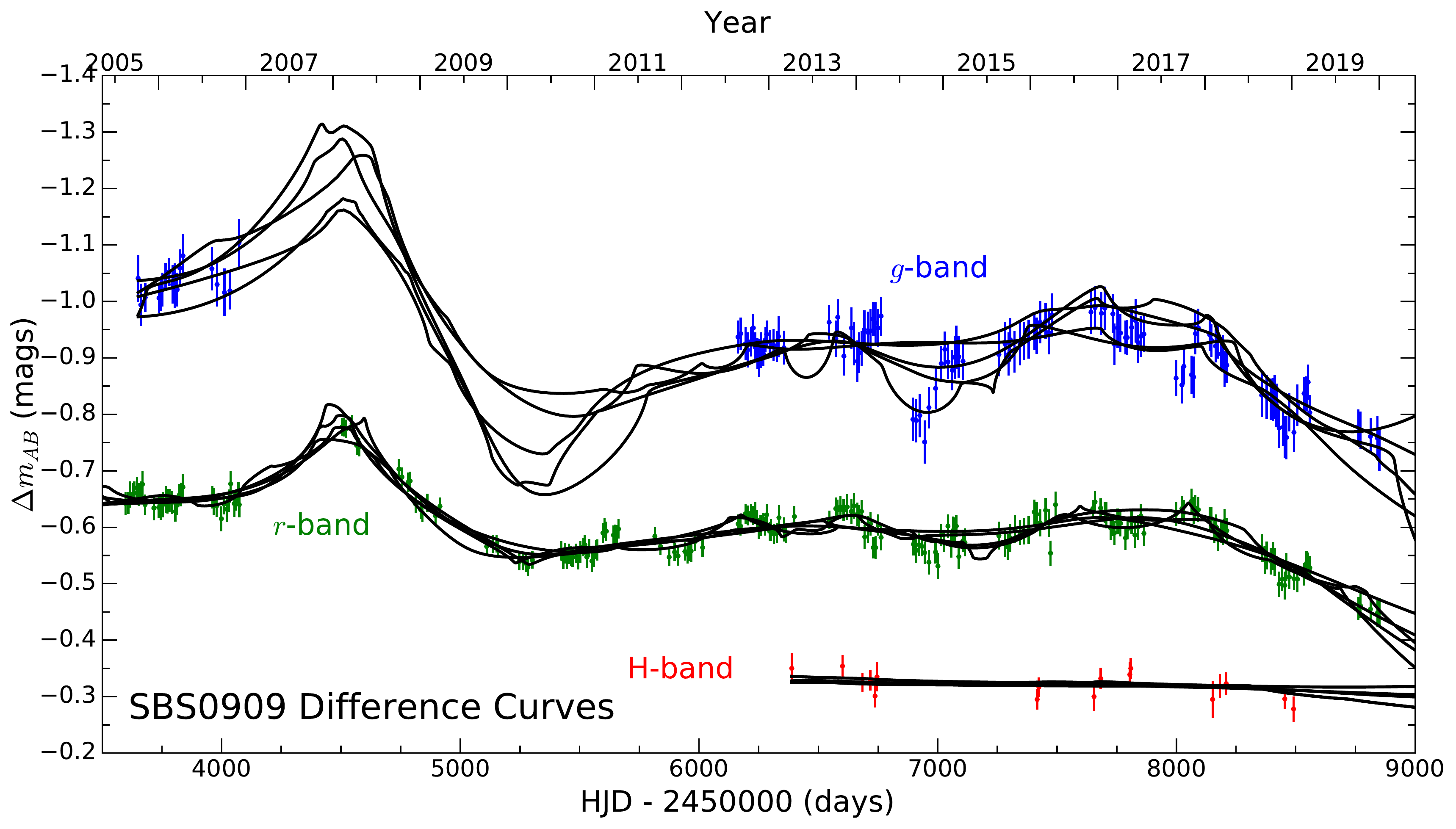}
\caption{The five best fit microlensing curves to the time--delay shifted difference curves for Q0957 (top) and SBS0909 (bottom).  The $g$--band fit is shown in blue, the $r$--band in green, and the $H$--band in red.  For the Q0957 $H$--band difference we do not show interpolated or extrapolated points with error bars $\geq 0.1$, although we do include these points in the joint analysis.  All difference light curves are vertically offset for clarity.}
\label{fig:bestfit_diff}
\end{figure*}

\begin{figure*}[t]
\centering
\plottwo{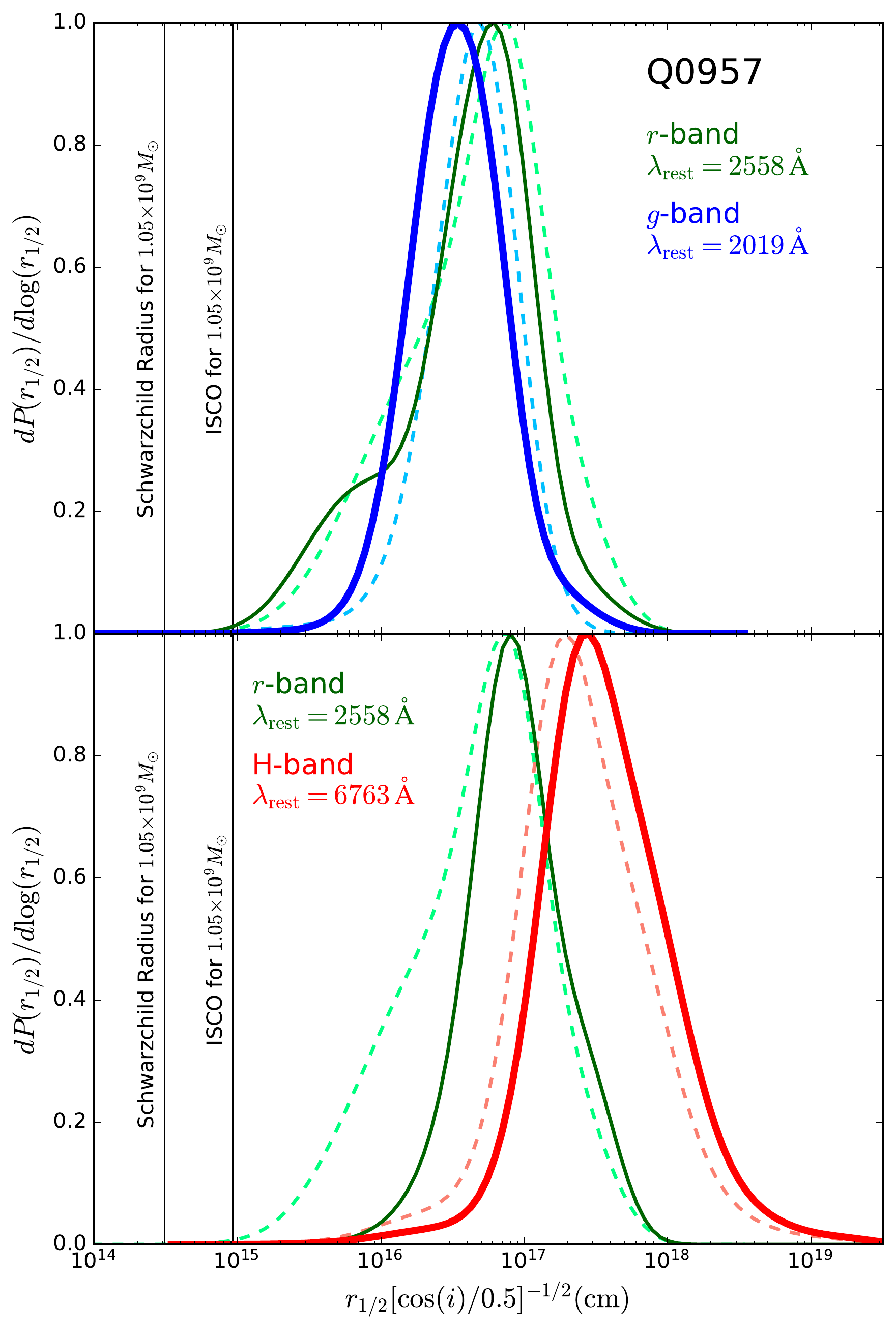}{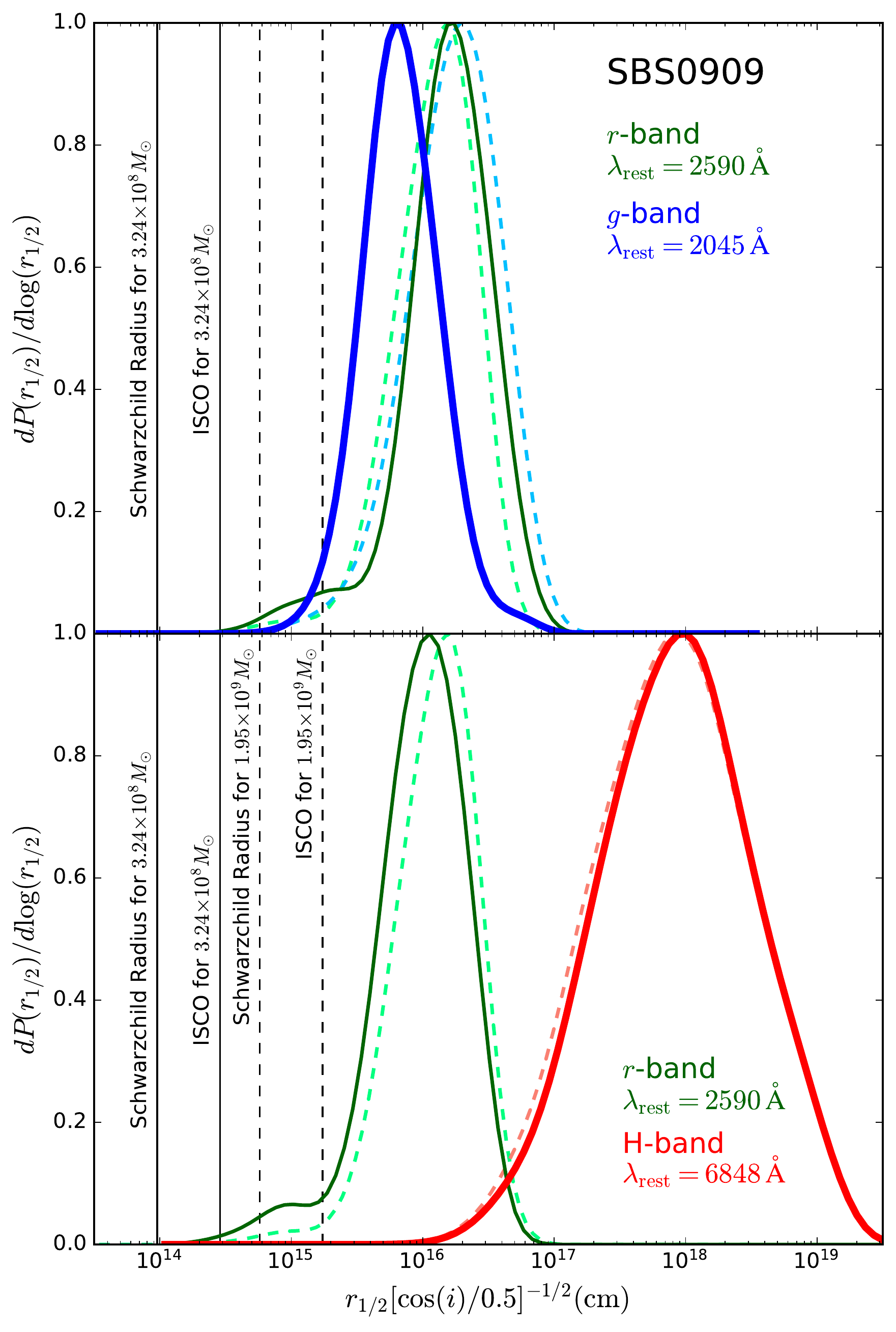}
\caption{Probability density of the size measurements in $r$--, $g$--, and $H$--bands for Q0957 (Left) and SBS0909 (Right).  Sizes are reported as a half light radius scaled to a $60\degr$ inclination angle.  Top: comparison between the $r$--band (green) and $g$--band (blue) sizes.  The solid lines ($g$--band is the heavier line) indicate the results of the joint analysis while the faint dashed lines show the independent analysis in each band.  Bottom: Comparison between the $r$--band (green) and $H$--band (red) sizes.  The solid green line gives the $r$--band under the joint analysis while the heavier solid red line gives joint analysis $r/H$--band results.  The faint dashed green line gives the $r$--band-only result while the faint dashed red line shows the constraint on $H$--band size if we include 20\% large scale flux dilution.
%\textbf{For Q0957 the dot-dashed orange line gives the $H$--band size found under the joint $g/H$--band analysis.}
The vertical lines show the Schwarzchild radius and innermost stable circular orbit (ISCO) for the black hole masses found by \citet{asse2011a}.  For SBS0909 we show lower and upper mass estimates of $10^{8.51}\,M_{\odot}$ using \ion{C}{4} (solid line) and $10^{9.29}\,M_{\odot}$ using H$\beta$ (dashed line).}
\label{fig:rspost}
\end{figure*}

\begin{figure*}[t]
\centering
\plottwo{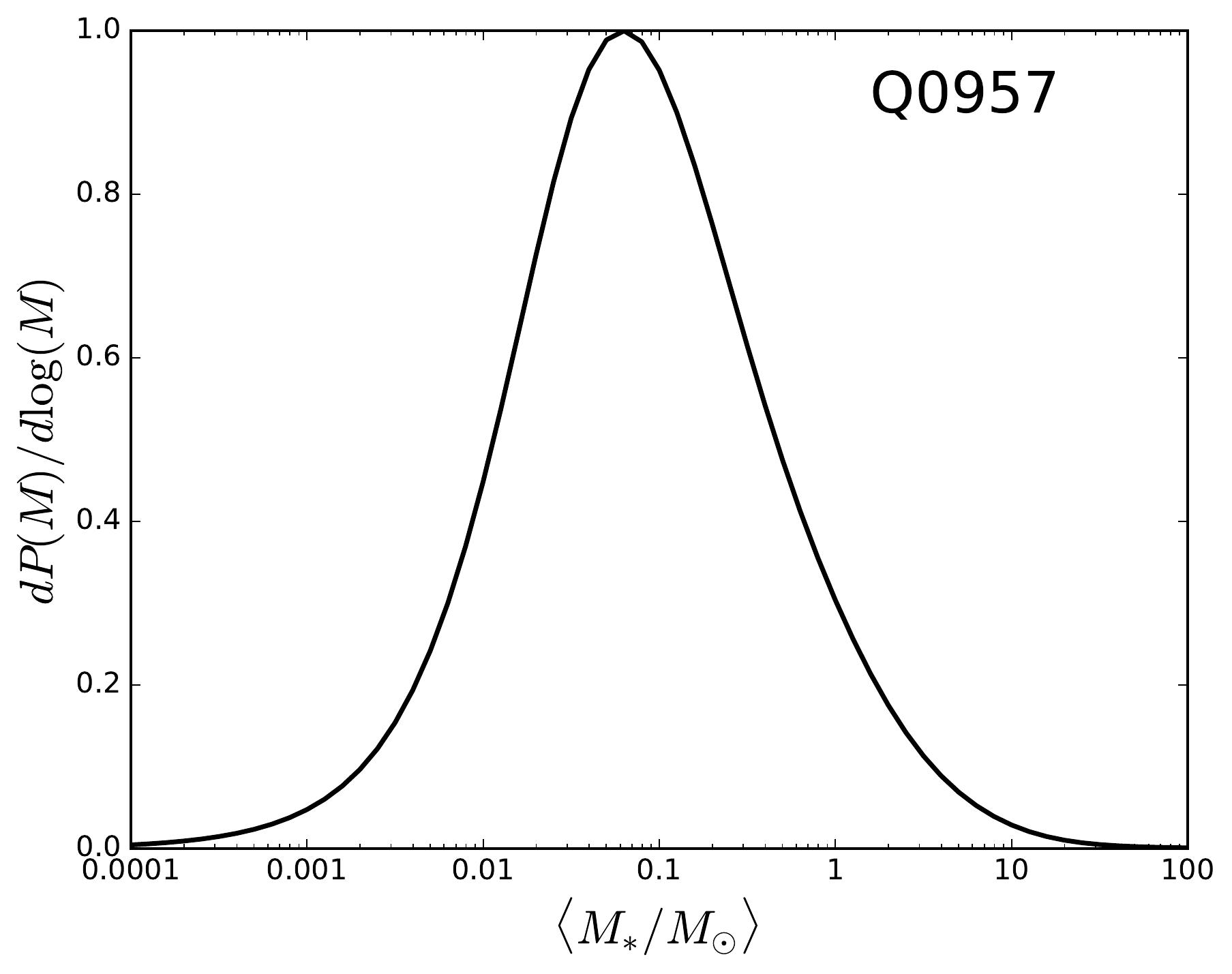}{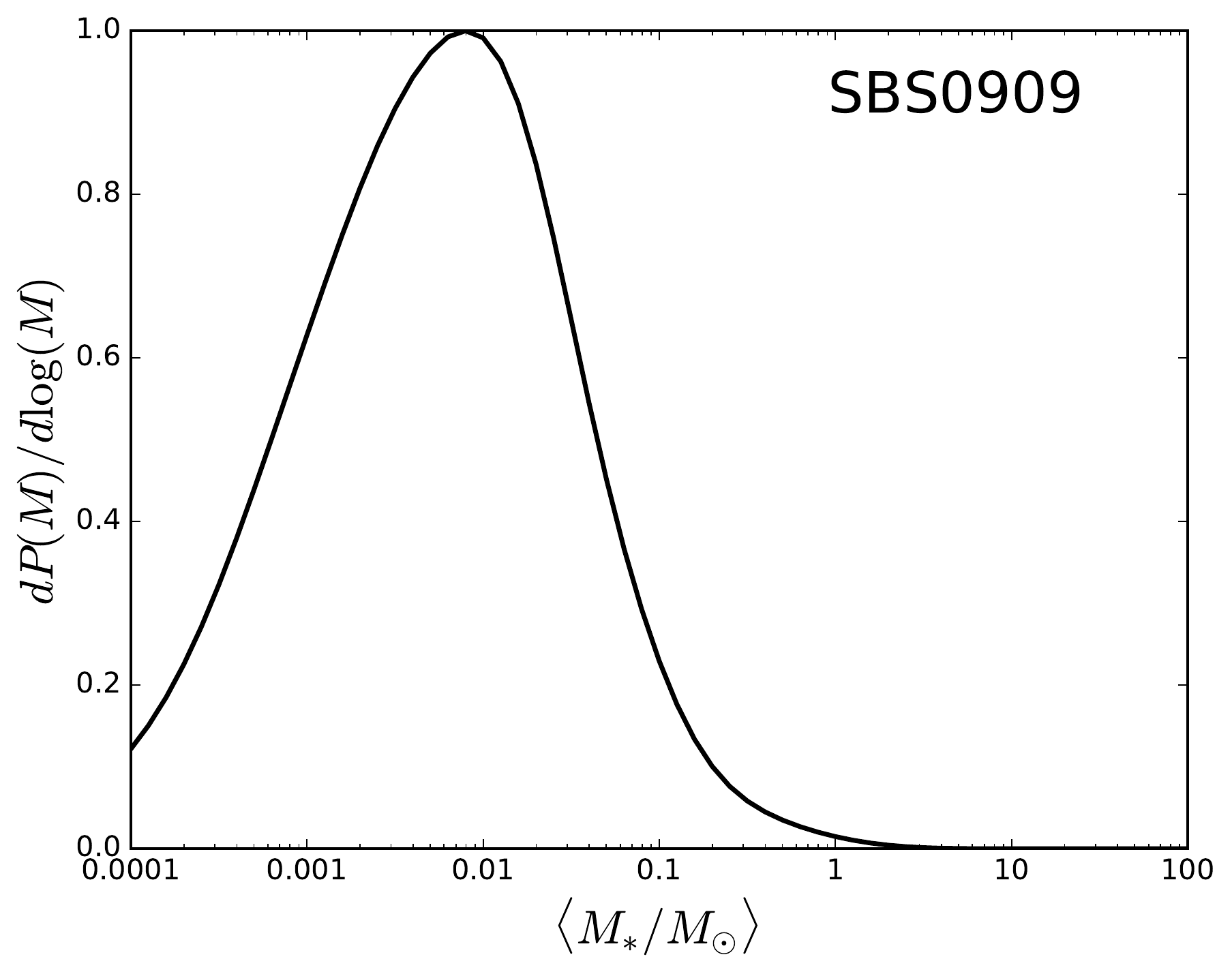}
\caption{Probability density of the mean microlens mass $\langle M_*/M_{\odot} \rangle$ in Q0957 (left) and SBS0909 (right).}
\label{fig:mean_mass}
\end{figure*}

\begin{deluxetable*}{llccl}
\tablecaption{Microlensing size measurements.  Sizes are reported as half-light radii scaled to $60\degr$ inclination.  Our adopted values from the joint analyses are indicated in bold. \label{tab:sizes}}

\tablehead{
\colhead{Quasar} & \colhead{filter} & \colhead{$\log${$(r_{1/2}[\cos(i)/0.5]^{-1/2}\,\text{cm}^{-1})$}} & \colhead{$\lambda_{\text{rest}}$} & \colhead{Analysis Method}}
\startdata
Q0957 & $g$ & $16.65^{+0.28}_{-0.32}$ & $2019\,\text{\AA}$ & $g$--band only \\
 & & $\bm{16.54^{+0.33}_{-0.33}}$ & $2019\,\text{\AA}$ & $r/g$--band joint \\
 & & $16.55^{+0.29}_{-0.31}$ & $2019\,\text{\AA}$ & $g/H$--band joint \\
\hline & $r$ & $16.73^{+0.42}_{-0.61}$ & $2558\,\text{\AA}$ & $r$--band only \\
 & & $\bm{16.66^{+0.37}_{-0.62}}$ & $2558\,\text{\AA}$ & $r/g$--band joint \\
 & & $16.93^{+0.37}_{-0.33}$ & $2558\,\text{\AA}$ & $r$/$H$--band joint \\
\hline & $H$ & $17.53^{+0.48}_{-0.39}$ & $6763\,\text{\AA}$ & $r$/$H$--band joint \\
 & & $17.21^{+0.37}_{-0.35}$ & $6763\,\text{\AA}$ & $g$/$H$--band joint \\
 & & $\bm{17.37^{+0.49}_{-0.40}}$ & $6763\,\text{\AA}$ & $r$/$H$--band w/ 20\% dilution \\
 & & $16.91^{+0.53}_{-0.48}$ & $6763\,\text{\AA}$ & $r$/$H$--band w/ 50\% dilution \\
 \hline
SBS0909 & $g$ & $16.25^{+0.33}_{-0.38}$ & $2045\,\text{\AA}$ & $g$--band only \\
 & & $\bm{15.83^{+0.30}_{-0.28}}$ & $2045\,\text{\AA}$ & $r/g$--band joint \\
\hline & $r$ & $16.11^{+0.28}_{-0.34}$ & $2590\,\text{\AA}$ & $r$--band only \\
 & & $\bm{16.21^{+0.30}_{-0.34}}$ & $2590\,\text{\AA}$ & $r/g$--band joint \\
 & & $16.00^{+0.31}_{-0.36}$ & $2590\,\text{\AA}$ & $r$/$H$--band joint \\
\hline & $H$ & $17.93^{+0.59}_{-0.61}$ & $6848\,\text{\AA}$ & $r$/$H$--band joint \\
 & & $\bm{17.90^{+0.61}_{-0.63}}$ & $6848\,\text{\AA}$ & $r$/$H$--band w/ 20\% dilution \\
 & & $17.83^{+0.66}_{-0.67}$ & $6848\,\text{\AA}$ & $r$/$H$--band w/ 50\% dilution \\
\enddata
\end{deluxetable*}

We show the five best-fit difference light curves from the $r$--band analysis and the $H$--band and $g$--band joint analysis in the top panel of Figure~\ref{fig:bestfit_diff} for Q0957.  Q0957 is undergoing a roughly linear microlensing event continuing nearly a decade since its original report in \citet{hain2012a}, observable both in $r$-- and $g$--band.  This places stronger constraints on the sizes than previously reported.  The sparse $H$--band difference curve is also plotted.

The Q0957 difference curves, especially in the $g$--band, display features around the 2007, 2009, and 2015 seasons that are related to the sharp variations of image A (see Figure~\ref{fig:lc_q0957}) and are not well-fit by the microlensing curves.  These features may be evidence of reprocessed broad line emission suggested in \citet{gilm2018a}, a hypothesis we plan to investigate in future work. After separately fitting subsections of the Q0957 light curves we find that failure to fit these features has a negligible impact on the microlensing sizes.
%sparsely sampled in this system, although we include interpolated and extrapolated points with large error bars as described in Section~\ref{sec:intrinsic}.  

For SBS0909 we show the difference curves in the bottom panel of Figure~\ref{fig:bestfit_diff}.  SBS0909 has continued to experience detectable microlensing variability since the size measurements reported in \citet{hain2013a}, better constraining both $r$-- and $g$--band sizes.  The $H$--band curve in SBS0909 is better sampled than that of Q0957, but shows minimal evidence of microlensing.

\subsection{Size Distributions}

From our microlensing analysis, we determine posterior distributions for the quasar continuum emission region sizes.  These are shown in Figure~\ref{fig:rspost} for Q0957 (left) and SBS0909 (right).  In the top panel the solid lines indicate the results from the joint $r$-- and $g$--band analysis, which are the best-constrained results.  The faint dashed lines indicate the single band analysis posterior for both $r$-- and $g$--bands.  For Q0957 the single band findings are consistent with the joint analysis though they favor a slightly larger size.
In SBS0909 the $r$--band size is tightly constrained and does not vary significantly with analysis method.  The $g$--band-only analysis, however, favors a larger continuum emission region size than the joint $g$--band analysis.  This is most likely because the $g$--band-only difference curve did not cover the significant microlensing event between 2007 and 2010 and fits to this event favor a smaller source size.  There are, however, relatively few good fits to the joint $r/g$--band curves (see Table~\ref{tab:nruns}), due to the very high microlensing variability in this system.

The bottom panel shows the $r$--band and $H$--band posterior size distributions, with joint constraints indicated with a solid line.  As in the top panel, the faint dashed green line indicates the $r$--band only analysis results and the solid red line indicates the $H$--band size from the joint $r/H$--band analysis.
The faint dashed red line indicates the $H$--band size measurements if we assume 20\% of the flux is emitted at large (unmicrolensed) scales.
%\textbf{In Q0957 the dot-dashed orange line show the joint $g/H$--band size measurements with 20\% large scale flux emission.}
%The faint dashed red line indicates the $H$--band size measurements with the commonly adopted median microlens mass prior of $0.1 < \langle M_*/M_{\odot} \rangle < 1$.
As expected from the sparse difference curve, the Q0957 $H$--band size measurement gives the broadest distribution, favoring a large $H$--band size.
%\textbf{The very high upper limit on the $H$--band size is a consequence of non-negligible probability for the combination of high $\hat{r}_s$ and low $\hat{v}_e$ solutions.
This distribution shifts noticeably as the flux dilution increases.  At 20\% flux dilution the $H$--band size estimate decreases by 31\%.  This size estimate with dilution is similar to the size found when jointly analyzing the $g/H$--band curves.  With 50\% flux dilution the $H$--band size becomes more comparable to the $r$--band size.  While the 50\% dilution case may not be physically likely, this shows that large scale flux emission can have a significant impact on size measurements.  In any case, the $H$--band sizes under varying dilution factors are formally consistent given the high uncertainty on each measurement.  

%However, we do find a robust lower size limit.  
%which is consistent with the size estimated when imposing the mass prior.
Despite a longer $H$--band light curve in SBS0909, we also find a broad distribution.  Furthermore, in SBS0909 the distribution of $H$--band values of $\hat{r}_s$ peaks at the size limit of our magnification patterns, $\approx 18.1\langle M/M_{\odot}\rangle^{1/2}\,\text{cm}$.  This means that we cannot explore larger sizes and must consider these $H$--band size measurements a robust lower limit only.  The distribution shifts toward smaller sizes for increasing flux dilution, but not as significantly as for Q0957.  Even with 50\% flux dilution the $H$--band lower limit is larger than the $r$--band size estimate.  

All size measurements are reported in Table~\ref{tab:sizes}, where the median posterior scale size is cast as a half light radius, $r_{1/2}$, and scaled to a nominal $60\degr$ inclination ($\cos(i) = 0.5$).
%The corresponding half-light radius, $r_{1/2} = 2.44r_s$, measurements are also provided.
We indicate the rest-frame wavelengths, based upon the redshifted central wavelengths for each band.  The bold results are those we adopt as our reported sizes.  These measurements come from our joint analyses, which are the most robustly constrained size measurements.  For the $H$--band we adopt the sizes under 20\% flux dilution as the most physically likely case.

%Our microlensing analysis also gives posterior distributions for the effective source velocity, the median microlens mass, and the convergence ratio $\kappa_*/\kappa$.  Our analysis returns strong constraints on these values and we provide plots of these distributions in the Appendix.

\subsection{Mean Microlens Mass}

Microlensing is uniquely poised to uncover information about the statistics of stars in the lens galaxy \citep{pool2019a}.  While we have not optimized our study to investigate lens galaxy stars, we naturally compute a distribution for the mean microlens mass (see Section~\ref{sec:ml_single}).  We show these distributions for both Q0957 and SBS0909 in Figure~\ref{fig:mean_mass}.

 Because the width of the mass distribution is proportional to the square of the width of the velocity distribution, these constraints are broad.  The Q0957 distribution favors relatively small mass stars and is consistent with the $0.52\,M_{\odot}$ microlens mass found in Q2237 \citep{point2010a}.  The SBS0909 distribution allows for a significant contribution from microlenses with low mass, including brown dwarfs and free-floating Jupiters.  Though constraints from microlensing in the Milky Way disfavor significant mass in these objects \citep{mroz2017a}, X-ray microlensing of quasars can be interpreted as evidence of an abundance of free-floating planets in lens galaxies \citep{bhat2019a}.  Degeneracies between the mean microlens mass and the unknown source velocity limit any conclusions we can draw here, but this may be an interesting avenue for further study.
 
\subsection{Size Scaling with Wavelength}
\label{sec:results_beta}

\begin{figure}[t]
\centering
\plotone{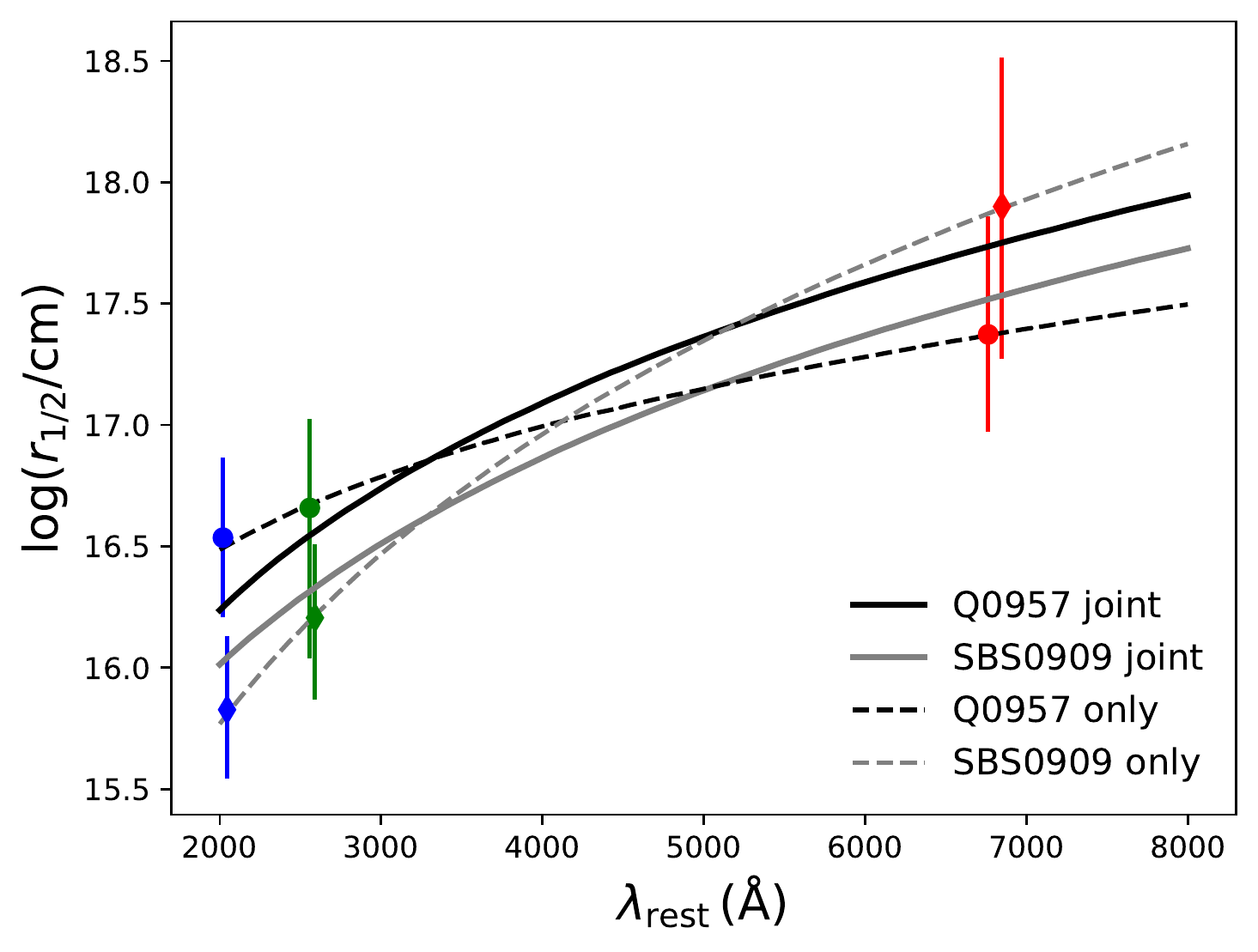}
\caption{Continuum emission region size scaling with wavelength, $r\propto \lambda^{1/\beta}$.  Microlensing size measurements for Q0957 are shown with solid circles and for SBS0909 with solid diamonds.  The $g$--, $r$--, and $H$--band sizes are indicated in blue, green, and red respectively.  Best-fit curves are shown for the two systems combined (solid lines) and the dashed lines for the individual systems where Q0957 is black and SBS0909 is gray.}
\label{fig:beta_curve}
\end{figure}

We use the sizes reported in Table~\ref{tab:sizes} to estimate the slope $\beta$ in $r\propto \lambda^{1/\beta}$ using a maximum likelihood fit in log-space to $\log(r_{1/2}) = (1/\beta)\log(\lambda) + C$.  We use the distributions shown in  Figure~\ref{fig:rspost} as the size priors in each band, using those associated with the bold lines in Table~\ref{tab:sizes}, and for our wavelength we use the central rest frame wavelength given in Table~\ref{tab:sizes}. We adopt the standard $\chi^2$ metric for our likelihood calculation.
%and assume a uniform prior on $\beta$ on the interval [0.1,2].  We also repeat for a Gaussian prior on $\beta$ with a mean value of the thin disk $\beta=0.75$ and a relatively broad width of $0.25$.
We run this analysis in each individual system and in the joint sample, where both quasars have a different constant, $C$, but the same scaling $\beta$.
%We also check the robustness of our solution by performing the analysis in both linear space and log space.

The best fit curves are shown in Figure~\ref{fig:beta_curve}.  In the joint analysis, using the $H$--band sizes from the $r/H$--band joint analyses with 20\% flux dilution, we find 
$\beta = 0.35^{+0.16}_{-0.08}$. Analyzing the individual systems separately we find $\beta = 0.54^{+0.58}_{-0.22}$ for Q0957 and $\beta = 0.25^{+0.13}_{-0.06}$ for SBS0909. % 20% dilution case
%$\beta = 0.28^{+0.12}_{-0.07}$. Analyzing the individual systems separately we find $\beta = 0.32^{+0.30}_{-0.12}$ for Q0957 and $\beta = 0.25^{+0.12}_{-0.06}$ for SBS0909. % 0 dilution case
The slope for Q0957, while formally consistent with the fiducial \citet{shak1973a} thin disk value of $\beta=3/4$, is suggestive of a shallower slope.  The slope for SBS0909 is much shallower but even so our joint analysis does not exclude the thin disk model at $2\sigma$ ($\beta=0.25^{+0.54}_{-0.13}$).  If we compute $\beta$ again increasing the $H$--band large--scale flux dilution to 50\%, we find a larger joint system value of $\beta = 0.42^{+0.25}_{-0.11}$.  The individual system values also increase to $\beta = 0.63^{+1.33}_{-1.90}$ for Q0957 and $\beta = 0.26^{+0.15}_{-0.07}$ for SBS0909.
We find a similar value of $\beta$ when using the distribution for the $H$--band size from the Q0957 joint $g/H$--band analysis.
We also attempted to fit the joint sample excluding the $H$--band, but the uncertainty in our findings was so high that $\beta$ was essentially unconstrained.

\section{Discussion}
\label{sec:discuss}

%We can compare our size measurements to the previous multi-epoch findings \citep{hain2012a, hain2013a} and those reported in single epoch microlensing analyses of Q0957 \citep{mott2012a, jime2014a} and SBS0909 \citep{medi2011a, jime2014a}.

For Q0957 our updated $r$--band size measurement is completely consistent with the findings of \citet{hain2012a}, though our constraints are now much tighter, as expected from a longer light curve with significant microlensing.
The reverberation-based size measurement of the $r$--band source in Q0957 is consistent with our microlensing-based size measurement, although it is close to the lower limit of our $1\,\sigma$ interval \citep{gilm2012a, gilm2018a}.  These measurements may show closer agreement if we include an analysis of broad line emission delays proposed in \citet{gilm2018a}.
Q0957 has also been studied using the single-epoch microlensing technique. \citet{jime2014a} analyzed time-delay corrected flux ratios from \citet{mott2012a} to yield size estimates at three wavelengths between $1216\,\text{\AA}$ to $2796\,\text{\AA}$.
%For single epoch microlensing measurements \citet{mott2012a} found Q0957 flux ratios from spectra collected by the Hubble Space Telescope Imaging Spectrograph (\textsl{HST}/STIS) in 1999 and 2000 (offset by the $420\,\text{day}$ time delay).  \citet{jime2014a} analyzed these flux ratios at three wavelengths between $1216\,\text{\AA}$ to $2796\,\text{\AA}$.
To compare to our findings, we converted their Gaussian scale radius to a half-light radius, scaled to our nominal $60\degr$ inclination, and computed sizes at the rest-frame wavelengths of each band, giving sizes of $\log(r_{1/2}/\text{cm}[\cos(i)/0.5]^{1/2}) = 16.55$ and $16.58$ in $g$--band and $r$--band respectively at a nominal $0.3\,M_{\odot}$ mean microlens mass.  Although their observations did not cover longer wavelengths, their fits predict an $H$--band size of $\log(r_{1/2}/\text{cm}[\cos(i)/0.5]^{1/2}) = 16.69$. These sizes are consistent with ours within formal statistical errors, although the \citet{jime2014a} findings suggest less chromatic variability.
%Our size measurements are most closely aligned if the \citet{jime2014a} results are scaled to a mean microlens mass of $0.35M_{\odot}$.

In SBS0909 both our $r$-- and $g$--band sizes are consistent with \citet{hain2013a}, although our reported sizes are larger.  This lessens the apparent tension between the \citet{hain2013a} size measurements and the large ISCO implied by the H$\beta$ measurement of the central black hole with a mass $10^{9.29}\, M_{\odot}$ \citep{asse2011a}.  Though our microlensing size measurements still disfavor such a large black hole, they are easily reconciled with the \ion{C}{4} mass estimate of $10^{8.51}\, M_{\odot}$ \citep{asse2011a}.  Microlensing signatures in merged spectra from this system were also analyzed by \citet{medi2011a} and re-analyzed (with the same data) by \citet{jime2014a}.  In both analyses the wavelength range spans all three of our bands. We again scale the reported results of both studies to half-light radii at $60\degr$ inclination, adopt a $0.3\,M_{\odot}$ mean microlens mass, and compute the sizes at the rest wavelengths of each of our bands.  For \citet{medi2011a} this gives sizes of $\log(r_{1/2}/\text{cm}[\cos(i)/0.5]^{1/2}) = 16.37, 16.47,$ and $16.85$ in $g$--band, $r$--band, and $H$--band respectively.  These findings are fully consistent with the size estimates from \citet{jime2014a} of $\log(r_{1/2}/\text{cm}[\cos(i)/0.5]^{1/2}) = 16.33, 16.42,$ and $16.78$ in $g$--band, $r$--band, and $H$--band respectively.  Though formally consistent with our findings, these sizes are systematically larger than our values reported in Table~\ref{tab:sizes}.  A smaller mean microlens mass, consistent with our distribution shown in Figure~\ref{fig:mean_mass}, would help reconcile these size measurements, though an abundance of such small mass objects in the lens galaxy may be physically unlikely.
%and are more closely aligned if the sizes are scaled from a $1M_{\odot}$ mean microlens mass to $0.05M_{\odot}$, which is consistent with the mean microlens mass distribution found in our analysis.

%\citet{jime2014a} found a very shallow temperature slope of $\beta\approx 2.0$.

As discussed in \citet{corn2020b}, a reliable measurement of the temperature profile slope, $\beta$, could help refine physical models of quasar accretion.  In this study we found evidence for a shallow profile of
$\beta = 0.35^{+0.16}_{-0.08}$, with individual estimates of $\beta = 0.54^{+0.58}_{-0.22}$ in Q0957 and $\beta = 0.25^{+0.13}_{-0.06}$ in SBS0909. % 20 percent flux dilution
%$\beta = 0.28^{+0.12}_{-0.07}$, with individual estimates of $\beta\approx0.32$ in Q0957 and $\beta\approx0.25$ in SBS0909. % No flux dilution
In stark contrast, \citet{jime2014a} found evidence for a steep temperature profiles of $\beta= 2.0$ in Q0957 and $\beta= 1.3$ in SBS0909.  \citet{medi2011a} also found a steeper temperature profile slope of $\beta= 1.1$ in SBS0909.
Although the error bars on each individual slope measurement are large, the disagreement is dramatic, particularly when comparing the 8 system aggregate value of $\beta=1.25\pm 0.2$ \citep{jime2014a} to our joint slope of 
$\beta = 0.35^{+0.16}_{-0.08}$, 
%$\beta = 0.28^{+0.12}_{-0.07}$,
findings which are formally inconsistent. The large difference between these temperature profiles has significant implications for the underlying structure of the quasar continuum emission region.

This disagreement is similar to that found by \citet{corn2020b} using a model-based approach with aggregated multi-epoch measurements at $\lambda_{\text{rest}}\approx 2500\,\text{\AA}$.  Both the cause for the contention and the path to a resolution with future measurements are not readily apparent.  One potential limitation of single epoch chromatic microlensing is that in the absence of strong chromatic variation, the single epoch method may favor an overly steep temperature profile \citep{bate2018a}.  The single epoch flux ratios for Q0957 are based on spectra collected in 1999 and 2000 \citep{mott2012a}, a period during which our difference light curves show no significant microlensing variability.  This could explain why Q0957 has such a large and poorly constrained temperature profile slope.  From our SBS0909 light curves it is less apparent whether the chromatic variation is low during the observations from \citet{medi2011a}, as some of these spectra predate our light curves.  The spectrum used in their analysis was, however, a composite of spectra collected over the span of several years so it is possible that intrinsic quasar variability influenced the single epoch results in this system.

Our measurement for the slope is preliminary and only uses a sample of two quasars.  Small changes in size measurements can have a large impact on the slope.  For example, the difference between our joint $g$--band size in SBS0909 and the $g$--band only analysis (see Table~\ref{tab:sizes}) would dramatically change our result.  With only 827 good fits, the joint $g$--band size in SBS0909 is the least robust of all our measurements.  Of the two systems in our study, SBS0909 has the shallower slope which may turn out to be, at least in part, impacted by this uncertainty.
%\textbf{We also found for Q0957 there are small systematic shifts in the inferred $\beta$ depending on whether shift in $H$--band size when using the joint $g$--band rather than the joint $r$--band analysis, brings the estimated slope into closer alignment with the thin disk profile.}
%Furthermore, the $H$--band measurements are at a rest frame wavelength of $\lambda_{\text{rest}}\approx6800\,\text{\AA}$ which is beyond the $5000\,\text{\AA}$ break found in \citet{vand2001a} quasar composite spectra.  Though the authors suggests some of this spectral break is due to host galaxy starlight, this break may be due in part to a change in the underlying spectral continuum shape.
Furthermore the $H$--band light curves likely contain a blend of the compact continuum emission
%, commonly associated with an accretion disk, 
and larger scale emission from host galaxy stars, reprocessed high energy emission, and the H$\alpha$ emission line.  When including a diluting flux fraction, we found that $H$--band sizes decrease, substantially in the case of Q0957, with the percentage of contaminating flux.  In Section~\ref{sec:results_beta}, we found that $\beta$ steepens when accounting for this dilution.  Though our measurements still favor a shallower value of $\beta \approx 0.42$ under 50\% flux dilution, this effect lessens the tension between our findings and those of other studies. While we know the Gaussian and thin disk source profiles return consistent sizes \citep{mort2005a}, we did not explore whether modeling the source as a thin disk affects the temperature profile slope.  This may be an important avenue for future study, though since \citet{medi2011a} and \citet{jime2014a} both used Gaussian sources, it is unlikely to resolve the discrepancy between studies.
%This adds uncertainty to the microlensing analysis and may explain in part why our $H$--band sizes are so large.  
%If indeed the $H$--band size associated with the compact continuum emission region is smaller, this will require a steeper value of $\beta$, lessening the tension between our findings and those of other studies.
Measurements including more wavelengths, particularly those between the $r$--band and $H$--band, may give the best constraints on $\beta$ and allow a better characterization of the underlying quasar continuum emission region structure.

%Quasar spectral slopes indicate a break at $5000\,\text{\AA}$ \citep{vand2001a} and models like that of \citet{jian2016a} predict a break at $\approx 100 r_g$, which corresponds roughly to $\lambda{\sim}2500\,\text{\AA}$.

 The results here show the utility of multi-epoch microlensing to measure quasar sizes in multiple bands spanning a wide range of quasar rest wavelengths, including $H$--band ground--based observations.  With only two quasars we see emerging evidence of a shallow temperature profile slope.  These constraints will be improved dramatically even with a modest sample size.  LSST \citep{lsst2009a} is developing a microlensing monitoring program which will produce multi-epoch light curves in $\mathcal{O}(1000)$ lensed quasars covering several observing bands in this key wavelength range. Microlensing events will be used as triggering for complementary single epoch microlensing measurements, which well help check for consistency between methods.  Combining the technique here with this expansive program may yield tremendous insights into quasar structure in the coming decade.

%Our sources of systematic error are unclear too.  The most likely source is contamination from emission lines [citations] in which the flux is emitted from different scales.  These can have an impact on inferred sizes, generally serving to reduce the overall sizes.  Indeed the findings of \citet{corn2020b} are most consistent with our findings here for large contamination from broad line or scattered light.  The effect this contamination will have on the temperature profile slope is still uncertain.

\acknowledgements
We thank Christopher Kochanek for the use of his microlensing analysis code.  We also thank Bonnie Lucas and the USNA ARC team for their help using TheARC computing cluster.

This material is based upon work supported by the National Science Foundation under grants AST-1614018 and AST-2007680 to M.A.C. and C.W.M.

%USNO wishes to acknowledge astronomer observing assistant Trudy Tilleman who obtained most of the observations for this paper, managed the resulting database, and made it available to USNA colleagues.

This paper is partially based on observations made with the Liverpool Telescope, which is operated on the island of La Palma by Liverpool John Moores University in the Spanish Observatorio del Roque de los Muchachos of the Instituto de Astrofisica de Canarias with financial support from the UK Science and Technology Facilities Council. We thank the staff of this robotic telescope for a kind interaction before, during, and after the observations. VNS and LJG have been supported by the
MINECO/AEI/FEDER-UE grant AYA2017-89815-P and University of Cantabria funds.

\facilities{USNO:61in, Liverpool:2m}
\software{Anaconda, \texttt{PyCS} \citep{tewe2013a}, IMFITFITS \citep{mcle1998a, leha2000a}, \texttt{astroscrappy} \citep[https://zenodo.org/record/1482019]{vand2001b}, \texttt{celerite}\citep{fore2017a}}
%PyCS, emcee, astroreduction (name...), 

\clearpage

\bibliographystyle{aasjournal}
\bibliography{usna_bibtex_archive}

\begin{thebibliography}{}
\expandafter\ifx\csname natexlab\endcsname\relax\def\natexlab#1{#1}\fi
\providecommand{\url}[1]{\href{#1}{#1}}
\providecommand{\dodoi}[1]{doi:~\href{http://doi.org/#1}{\nolinkurl{#1}}}
\providecommand{\doeprint}[1]{\href{http://ascl.net/#1}{\nolinkurl{http://ascl.net/#1}}}
\providecommand{\doarXiv}[1]{\href{https://arxiv.org/abs/#1}{\nolinkurl{https://arxiv.org/abs/#1}}}

\bibitem[{{Antonucci}(2013)}]{anto2013a}
{Antonucci}, R. 2013, \nat, 495, 165, \dodoi{10.1038/495165a}

\bibitem[{{Assef} {et~al.}(2011){Assef}, {Denney}, {Kochanek}, {Peterson},
  {Koz{\l}owski}, {Ageorges}, {Barrows}, {Buschkamp}, {Dietrich}, {Falco},
  {Feiz}, {Gemperlein}, {Germeroth}, {Grier}, {Hofmann}, {Juette}, {Khan},
  {Kilic}, {Knierim}, {Laun}, {Lederer}, {Lehmitz}, {Lenzen}, {Mall}, {Madsen},
  {Mandel}, {Martini}, {Mathur}, {Mogren}, {Mueller}, {Naranjo}, {Pasquali},
  {Polsterer}, {Pogge}, {Quirrenbach}, {Seifert}, {Stern}, {Shappee}, {Storz},
  {Van Saders}, {Weiser}, \& {Zhang}}]{asse2011a}
{Assef}, R.~J., {Denney}, K.~D., {Kochanek}, C.~S., {et~al.} 2011, \apj, 742,
  93, \dodoi{10.1088/0004-637X/742/2/93}

\bibitem[{{Bate} {et~al.}(2008){Bate}, {Floyd}, {Webster}, \&
  {Wyithe}}]{bate2008a}
{Bate}, N.~F., {Floyd}, D.~J.~E., {Webster}, R.~L., \& {Wyithe}, J.~S.~B. 2008,
  \mnras, 391, 1955, \dodoi{10.1111/j.1365-2966.2008.14020.x}

\bibitem[{{Bate} {et~al.}(2018){Bate}, {Vernardos}, {O'Dowd}, {Neri-Larios},
  {Webster}, {Floyd}, {Barone- Nugent}, {Labrie}, {King}, \&
  {Yong}}]{bate2018a}
{Bate}, N.~F., {Vernardos}, G., {O'Dowd}, M.~J., {et~al.} 2018, \mnras, 479,
  4796, \dodoi{10.1093/mnras/sty1793}

\bibitem[{{Bentz} {et~al.}(2010){Bentz}, {Walsh}, {Barth}, {Yoshii}, {Woo},
  {Wang}, {Treu}, {Thornton}, {Street}, {Steele}, {Silverman}, {Serduke},
  {Sakata}, {Minezaki}, {Malkan}, {Li}, {Lee}, {Hiner}, {Hidas}, {Greene},
  {Gates}, {Ganeshalingam}, {Filippenko}, {Canalizo}, {Bennert}, \&
  {Baliber}}]{bent2010a}
{Bentz}, M.~C., {Walsh}, J.~L., {Barth}, A.~J., {et~al.} 2010, \apj, 716, 993,
  \dodoi{10.1088/0004-637X/716/2/993}

\bibitem[{{Bhatiani} {et~al.}(2019){Bhatiani}, {Dai}, \& {Guerras}}]{bhat2019a}
{Bhatiani}, S., {Dai}, X., \& {Guerras}, E. 2019, \apj, 885, 77,
  \dodoi{10.3847/1538-4357/ab46ac}

\bibitem[{{Birrer} {et~al.}(2019){Birrer}, {Treu}, {Rusu}, {Bonvin},
  {Fassnacht}, {Chan}, {Agnello}, {Shajib}, {Chen}, {Auger}, {Courbin},
  {Hilbert}, {Sluse}, {Suyu}, {Wong}, {Marshall}, {Lemaux}, \&
  {Meylan}}]{birr2019a}
{Birrer}, S., {Treu}, T., {Rusu}, C.~E., {et~al.} 2019, \mnras, 484, 4726,
  \dodoi{10.1093/mnras/stz200}

\bibitem[{{Blackburne} {et~al.}(2011){Blackburne}, {Pooley}, {Rappaport}, \&
  {Schechter}}]{blac2011a}
{Blackburne}, J.~A., {Pooley}, D., {Rappaport}, S., \& {Schechter}, P.~L. 2011,
  \apj, 729, 34, \dodoi{10.1088/0004-637X/729/1/34}

\bibitem[{{Blaes}(2004)}]{blae2004a}
{Blaes}, O.~M. 2004, in Accretion Discs, Jets and High Energy Phenomena in
  Astrophysics, ed. V.~{Beskin}, G.~{Henri}, F.~{Menard}, \& {et al.}, Vol.~78,
  137--185.
\newblock \doarXiv{astro-ph/0211368}

\bibitem[{{Bonning} {et~al.}(2013){Bonning}, {Shields}, {Stevens}, \&
  {Salviander}}]{bonn2013a}
{Bonning}, E.~W., {Shields}, G.~A., {Stevens}, A.~C., \& {Salviander}, S. 2013,
  \apj, 770, 30, \dodoi{10.1088/0004-637X/770/1/30}

\bibitem[{{Bonvin} {et~al.}(2017){Bonvin}, {Courbin}, {Suyu}, {Marshall},
  {Rusu}, {Sluse}, {Tewes}, {Wong}, {Collett}, {Fassnacht}, {Treu}, {Auger},
  {Hilbert}, {Koopmans}, {Meylan}, {Rumbaugh}, {Sonnenfeld}, \&
  {Spiniello}}]{bonv2017a}
{Bonvin}, V., {Courbin}, F., {Suyu}, S.~H., {et~al.} 2017, \mnras, 465, 4914,
  \dodoi{10.1093/mnras/stw3006}

\bibitem[{{Boyle} \& {Terlevich}(1998)}]{boyl1998a}
{Boyle}, B.~J., \& {Terlevich}, R.~J. 1998, \mnras, 293, L49,
  \dodoi{10.1046/j.1365-8711.1998.01264.x}

\bibitem[{{Cackett} {et~al.}(2018){Cackett}, {Chiang}, {McHardy}, {Edelson},
  {Goad}, {Horne}, \& {Korista}}]{cack2018a}
{Cackett}, E.~M., {Chiang}, C.-Y., {McHardy}, I., {et~al.} 2018, \apj, 857, 53,
  \dodoi{10.3847/1538-4357/aab4f7}

\bibitem[{{Cackett} {et~al.}(2007){Cackett}, {Horne}, \& {Winkler}}]{cack2007a}
{Cackett}, E.~M., {Horne}, K., \& {Winkler}, H. 2007, \mnras, 380, 669,
  \dodoi{10.1111/j.1365-2966.2007.12098.x}

\bibitem[{{Calderone} {et~al.}(2017){Calderone}, {Nicastro}, {Ghisellini},
  {Dotti}, {Sbarrato}, {Shankar}, \& {Colpi}}]{cald2017a}
{Calderone}, G., {Nicastro}, L., {Ghisellini}, G., {et~al.} 2017, \mnras, 472,
  4051, \dodoi{10.1093/mnras/stx2239}

\bibitem[{{Chang} \& {Refsdal}(1979)}]{chan1979a}
{Chang}, K., \& {Refsdal}, S. 1979, \nat, 282, 561, \dodoi{10.1038/282561a0}

\bibitem[{{Chartas} {et~al.}(2017){Chartas}, {Krawczynski}, {Zalesky},
  {Kochanek}, {Dai}, {Morgan}, \& {Mosquera}}]{char2017a}
{Chartas}, G., {Krawczynski}, H., {Zalesky}, L., {et~al.} 2017, \apj, 837, 26,
  \dodoi{10.3847/1538-4357/aa5d50}

\bibitem[{{Chen} {et~al.}(2019){Chen}, {Fassnacht}, {Suyu}, {Rusu}, {Chan},
  {Wong}, {Auger}, {Hilbert}, {Bonvin}, {Birrer}, {Millon}, {Koopmans},
  {Lagattuta}, {McKean}, {Vegetti}, {Courbin}, {Ding}, {Halkola}, {Jee},
  {Shajib}, {Sluse}, {Sonnenfeld}, \& {Treu}}]{chen2019a}
{Chen}, G. C.~F., {Fassnacht}, C.~D., {Suyu}, S.~H., {et~al.} 2019, \mnras,
  490, 1743, \dodoi{10.1093/mnras/stz2547}

\bibitem[{{Congdon} {et~al.}(2007){Congdon}, {Keeton}, \& {Osmer}}]{cong2007a}
{Congdon}, A.~B., {Keeton}, C.~R., \& {Osmer}, S.~J. 2007, \mnras, 376, 263,
  \dodoi{10.1111/j.1365-2966.2007.11426.x}

\bibitem[{{Cornachione} \& {Morgan}(2020)}]{corn2020b}
{Cornachione}, M.~A., \& {Morgan}, C.~W. 2020, \apj, 895, 93,
  \dodoi{10.3847/1538-4357/ab8aed}

\bibitem[{{Cornachione} {et~al.}(2020){Cornachione}, {Morgan}, {Millon},
  {Bentz}, {Courbin}, {Bonvin}, \& {Falco}}]{corn2020a}
{Cornachione}, M.~A., {Morgan}, C.~W., {Millon}, M., {et~al.} 2020, \apj, 895,
  125, \dodoi{10.3847/1538-4357/ab557a}

\bibitem[{{Dai} {et~al.}(2010){Dai}, {Kochanek}, {Chartas}, {Koz{\l}owski},
  {Morgan}, {Garmire}, \& {Agol}}]{dai2010a}
{Dai}, X., {Kochanek}, C.~S., {Chartas}, G., {et~al.} 2010, \apj, 709, 278,
  \dodoi{10.1088/0004-637X/709/1/278}

\bibitem[{{Dai} {et~al.}(2019){Dai}, {Steele}, {Guerras}, {Morgan}, \&
  {Chen}}]{dai2019a}
{Dai}, X., {Steele}, S., {Guerras}, E., {Morgan}, C.~W., \& {Chen}, B. 2019,
  \apj, 879, 35, \dodoi{10.3847/1538-4357/ab1d56}

\bibitem[{{Davis} {et~al.}(2007){Davis}, {Woo}, \& {Blaes}}]{davi2007a}
{Davis}, S.~W., {Woo}, J.-H., \& {Blaes}, O.~M. 2007, \apj, 668, 682,
  \dodoi{10.1086/521393}

\bibitem[{{Di Matteo} {et~al.}(2005){Di Matteo}, {Springel}, \&
  {Hernquist}}]{dima2005a}
{Di Matteo}, T., {Springel}, V., \& {Hernquist}, L. 2005, \nat, 433, 604,
  \dodoi{10.1038/nature03335}

\bibitem[{{Edelson} {et~al.}(2015){Edelson}, {Gelbord}, {Horne}, {McHardy},
  {Peterson}, {Ar{\'e}valo}, {Breeveld}, {De Rosa}, {Evans}, {Goad}, {Kriss},
  {Brandt}, {Gehrels}, {Grupe}, {Kennea}, {Kochanek}, {Nousek}, {Papadakis},
  {Siegel}, {Starkey}, {Uttley}, {Vaughan}, {Young}, {Barth}, {Bentz},
  {Brewer}, {Crenshaw}, {Dalla Bont{\`a}}, {De Lorenzo-C{\'a}ceres}, {Denney},
  {Dietrich}, {Ely}, {Fausnaugh}, {Grier}, {Hall}, {Kaastra}, {Kelly},
  {Korista}, {Lira}, {Mathur}, {Netzer}, {Pancoast}, {Pei}, {Pogge},
  {Schimoia}, {Treu}, {Vestergaard}, {Villforth}, {Yan}, \& {Zu}}]{edel2015a}
{Edelson}, R., {Gelbord}, J.~M., {Horne}, K., {et~al.} 2015, \apj, 806, 129,
  \dodoi{10.1088/0004-637X/806/1/129}

\bibitem[{{Edelson} {et~al.}(2019){Edelson}, {Gelbord}, {Cackett}, {Peterson},
  {Horne}, {Barth}, {Starkey}, {Bentz}, {Brandt}, {Goad}, {Joner}, {Korista},
  {Netzer}, {Page}, {Uttley}, {Vaughan}, {Breeveld}, {Cenko}, {Done}, {Evans},
  {Fausnaugh}, {Ferland}, {Gonzalez-Buitrago}, {Gropp}, {Grupe}, {Kaastra},
  {Kennea}, {Kriss}, {Mathur}, {Mehdipour}, {Mudd}, {Nousek}, {Schmidt},
  {Vestergaard}, \& {Villforth}}]{edel2019a}
{Edelson}, R., {Gelbord}, J., {Cackett}, E., {et~al.} 2019, \apj, 870, 123,
  \dodoi{10.3847/1538-4357/aaf3b4}

\bibitem[{{Eigenbrod} {et~al.}(2008){Eigenbrod}, {Courbin}, {Meylan}, {Agol},
  {Anguita}, {Schmidt}, \& {Wambsganss}}]{eige2008a}
{Eigenbrod}, A., {Courbin}, F., {Meylan}, G., {et~al.} 2008, \aap, 490, 933,
  \dodoi{10.1051/0004-6361:200810729}

\bibitem[{{Eulaers} \& {Magain}(2011)}]{eula2011a}
{Eulaers}, E., \& {Magain}, P. 2011, \aap, 536, A44,
  \dodoi{10.1051/0004-6361/201016101}

\bibitem[{{Fabian} {et~al.}(2000){Fabian}, {Iwasawa}, {Reynolds}, \&
  {Young}}]{fabi2000a}
{Fabian}, A.~C., {Iwasawa}, K., {Reynolds}, C.~S., \& {Young}, A.~J. 2000,
  \pasp, 112, 1145, \dodoi{10.1086/316610}

\bibitem[{{Fadely} {et~al.}(2010){Fadely}, {Keeton}, {Nakajima}, \&
  {Bernstein}}]{fade2010a}
{Fadely}, R., {Keeton}, C.~R., {Nakajima}, R., \& {Bernstein}, G.~M. 2010,
  \apj, 711, 246, \dodoi{10.1088/0004-637X/711/1/246}

\bibitem[{{Fischer} {et~al.}(2003){Fischer}, {Vrba}, {Toomey}, {Lucke}, {Wang},
  {Henden}, {Robichaud}, {Onaka}, {Hicks}, {Harris}, {Stahlberger},
  {Kosakowski}, {Dudley}, \& {Johnston}}]{fisc2003a}
{Fischer}, J., {Vrba}, F.~J., {Toomey}, D.~W., {et~al.} 2003, Society of
  Photo-Optical Instrumentation Engineers (SPIE) Conference Series, Vol. 4841,
  {ASTROCAM: offner re-imaging 1024 X 1024 InSb camera for near-infrared
  astrometry on the USNO 1.55-m telescope}, ed. M.~{Iye} \& A.~F.~M.
  {Moorwood}, 564--577, \dodoi{10.1117/12.461033}

\bibitem[{{Floyd} {et~al.}(2009){Floyd}, {Bate}, \& {Webster}}]{floy2009a}
{Floyd}, D. J.~E., {Bate}, N.~F., \& {Webster}, R.~L. 2009, \mnras, 398, 233,
  \dodoi{10.1111/j.1365-2966.2009.15045.x}

\bibitem[{{Foreman-Mackey} {et~al.}(2017){Foreman-Mackey}, {Agol},
  {Ambikasaran}, \& {Angus}}]{fore2017a}
{Foreman-Mackey}, D., {Agol}, E., {Ambikasaran}, S., \& {Angus}, R. 2017, \aj,
  154, 220, \dodoi{10.3847/1538-3881/aa9332}

\bibitem[{{Gil-Merino} {et~al.}(2012){Gil-Merino}, {Goicoechea}, {Shalyapin},
  \& {Braga}}]{gilm2012a}
{Gil-Merino}, R., {Goicoechea}, L.~J., {Shalyapin}, V.~N., \& {Braga}, V.~F.
  2012, \apj, 744, 47, \dodoi{10.1088/0004-637X/744/1/47}

\bibitem[{{Gil-Merino} {et~al.}(2018){Gil-Merino}, {Goicoechea}, {Shalyapin},
  \& {Oscoz}}]{gilm2018a}
{Gil-Merino}, R., {Goicoechea}, L.~J., {Shalyapin}, V.~N., \& {Oscoz}, A. 2018,
  \aap, 616, A118, \dodoi{10.1051/0004-6361/201832737}

\bibitem[{{Glikman} {et~al.}(2006){Glikman}, {Helfand}, \& {White}}]{glik2006a}
{Glikman}, E., {Helfand}, D.~J., \& {White}, R.~L. 2006, \apj, 640, 579,
  \dodoi{10.1086/500098}

\bibitem[{{Goicoechea} {et~al.}(2008){Goicoechea}, {Shalyapin}, {Koptelova},
  {Gil-Merino}, {Zheleznyak}, \& {Ull{\'a}n}}]{goic2008a}
{Goicoechea}, L.~J., {Shalyapin}, V.~N., {Koptelova}, E., {et~al.} 2008, \na,
  13, 182, \dodoi{10.1016/j.newast.2007.08.006}

\bibitem[{{Goicoechea} {et~al.}(2020){Goicoechea}, {Artamonov}, {Shalyapin},
  {Sergeyev}, {Burkhonov}, {Akhunov}, {Asfand iyarov}, {Bruevich},
  {Ehgamberdiev}, {Shimanovskaya}, \& {Zheleznyak}}]{goic2020a}
{Goicoechea}, L.~J., {Artamonov}, B.~P., {Shalyapin}, V.~N., {et~al.} 2020,
  \aap, 637, A89, \dodoi{10.1051/0004-6361/202037902}

\bibitem[{{Gould}(2000)}]{goul2000a}
{Gould}, A. 2000, \apj, 535, 928, \dodoi{10.1086/308865}

\bibitem[{{Hainline} {et~al.}(2012){Hainline}, {Morgan}, {Beach}, {Kochanek},
  {Harris}, {Tilleman}, {Fadely}, {Falco}, \& {Le}}]{hain2012a}
{Hainline}, L.~J., {Morgan}, C.~W., {Beach}, J.~N., {et~al.} 2012, \apj, 744,
  104, \dodoi{10.1088/0004-637X/744/2/104}

\bibitem[{{Hainline} {et~al.}(2013){Hainline}, {Morgan}, {MacLeod}, {Landaal},
  {Kochanek}, {Harris}, {Tilleman}, {Goicoechea}, {Shalyapin}, \&
  {Falco}}]{hain2013a}
{Hainline}, L.~J., {Morgan}, C.~W., {MacLeod}, C.~L., {et~al.} 2013, \apj, 774,
  69, \dodoi{10.1088/0004-637X/774/1/69}

\bibitem[{{Hern{\'a}n-Caballero} {et~al.}(2016){Hern{\'a}n-Caballero},
  {Hatziminaoglou}, {Alonso-Herrero}, \& {Mateos}}]{hern2016a}
{Hern{\'a}n-Caballero}, A., {Hatziminaoglou}, E., {Alonso-Herrero}, A., \&
  {Mateos}, S. 2016, \mnras, 463, 2064, \dodoi{10.1093/mnras/stw2107}

\bibitem[{{Hinshaw} {et~al.}(2009){Hinshaw}, {Weiland}, {Hill}, {Odegard},
  {Larson}, {Bennett}, {Dunkley}, {Gold}, {Greason}, {Jarosik}, {Komatsu},
  {Nolta}, {Page}, {Spergel}, {Wollack}, {Halpern}, {Kogut}, {Limon}, {Meyer},
  {Tucker}, \& {Wright}}]{hins2009a}
{Hinshaw}, G., {Weiland}, J.~L., {Hill}, R.~S., {et~al.} 2009, The
  Astrophysical Journal Supplement Series, 180, 225,
  \dodoi{10.1088/0067-0049/180/2/225}

\bibitem[{{Jiang} {et~al.}(2016){Jiang}, {Davis}, \& {Stone}}]{jian2016a}
{Jiang}, Y.-F., {Davis}, S.~W., \& {Stone}, J.~M. 2016, \apj, 827, 10,
  \dodoi{10.3847/0004-637X/827/1/10}

\bibitem[{{Jim{\'e}nez-Vicente} {et~al.}(2014){Jim{\'e}nez-Vicente},
  {Mediavilla}, {Kochanek}, {Mu{\~n}oz}, {Motta}, {Falco}, \&
  {Mosquera}}]{jime2014a}
{Jim{\'e}nez-Vicente}, J., {Mediavilla}, E., {Kochanek}, C.~S., {et~al.} 2014,
  \apj, 783, 47, \dodoi{10.1088/0004-637X/783/1/47}

\bibitem[{{Jim{\'e}nez-Vicente} {et~al.}(2012){Jim{\'e}nez-Vicente},
  {Mediavilla}, {Mu{\~n}oz}, \& {Kochanek}}]{jime2012a}
{Jim{\'e}nez-Vicente}, J., {Mediavilla}, E., {Mu{\~n}oz}, J.~A., \& {Kochanek},
  C.~S. 2012, \apj, 751, 106, \dodoi{10.1088/0004-637X/751/2/106}

\bibitem[{{Kochanek}(2004)}]{koch2004a}
{Kochanek}, C.~S. 2004, \apj, 605, 58, \dodoi{10.1086/382180}

\bibitem[{{Kochanek} {et~al.}(2006){Kochanek}, {Morgan}, {Falco}, {McLeod},
  {Winn}, {Dembicky}, \& {Ketzeback}}]{koch2006a}
{Kochanek}, C.~S., {Morgan}, N.~D., {Falco}, E.~E., {et~al.} 2006, \apj, 640,
  47, \dodoi{10.1086/499766}

\bibitem[{{Leh{\'a}r} {et~al.}(2000){Leh{\'a}r}, {Falco}, {Kochanek}, {McLeod},
  {Mu{\~n}oz}, {Impey}, {Rix}, {Keeton}, \& {Peng}}]{leha2000a}
{Leh{\'a}r}, J., {Falco}, E.~E., {Kochanek}, C.~S., {et~al.} 2000, \apj, 536,
  584, \dodoi{10.1086/308963}

\bibitem[{{LSST Science Collaboration} {et~al.}(2009){LSST Science
  Collaboration}, {Abell}, {Allison}, {Anderson}, {Andrew}, {Angel}, {Armus},
  {Arnett}, {Asztalos}, {Axelrod}, {Bailey}, {Ballantyne}, {Bankert},
  {Barkhouse}, {Barr}, {Barrientos}, {Barth}, {Bartlett}, {Becker}, {Becla},
  {Beers}, {Bernstein}, {Biswas}, {Blanton}, {Bloom}, {Bochanski}, {Boeshaar},
  {Borne}, {Bradac}, {Brandt}, {Bridge}, {Brown}, {Brunner}, {Bullock},
  {Burgasser}, {Burge}, {Burke}, {Cargile}, {Chand rasekharan}, {Chartas},
  {Chesley}, {Chu}, {Cinabro}, {Claire}, {Claver}, {Clowe}, {Connolly}, {Cook},
  {Cooke}, {Cooray}, {Covey}, {Culliton}, {de Jong}, {de Vries}, {Debattista},
  {Delgado}, {Dell'Antonio}, {Dhital}, {Di Stefano}, {Dickinson}, {Dilday},
  {Djorgovski}, {Dobler}, {Donalek}, {Dubois-Felsmann}, {Durech},
  {Eliasdottir}, {Eracleous}, {Eyer}, {Falco}, {Fan}, {Fassnacht}, {Ferguson},
  {Fernandez}, {Fields}, {Finkbeiner}, {Figueroa}, {Fox}, {Francke}, {Frank},
  {Frieman}, {Fromenteau}, {Furqan}, {Galaz}, {Gal-Yam}, {Garnavich},
  {Gawiser}, {Geary}, {Gee}, {Gibson}, {Gilmore}, {Grace}, {Green}, {Gressler},
  {Grillmair}, {Habib}, {Haggerty}, {Hamuy}, {Harris}, {Hawley}, {Heavens},
  {Hebb}, {Henry}, {Hileman}, {Hilton}, {Hoadley}, {Holberg}, {Holman},
  {Howell}, {Infante}, {Ivezic}, {Jacoby}, {Jain}, {R}, {Jedicke}, {Jee},
  {Garrett Jernigan}, {Jha}, {Johnston}, {Jones}, {Juric}, {Kaasalainen},
  {Styliani}, {Kafka}, {Kahn}, {Kaib}, {Kalirai}, {Kantor}, {Kasliwal},
  {Keeton}, {Kessler}, {Knezevic}, {Kowalski}, {Krabbendam}, {Krughoff},
  {Kulkarni}, {Kuhlman}, {Lacy}, {Lepine}, {Liang}, {Lien}, {Lira}, {Long},
  {Lorenz}, {Lotz}, {Lupton}, {Lutz}, {Macri}, {Mahabal}, {Mandelbaum},
  {Marshall}, {May}, {McGehee}, {Meadows}, {Meert}, {Milani}, {Miller},
  {Miller}, {Mills}, {Minniti}, {Monet}, {Mukadam}, {Nakar}, {Neill}, {Newman},
  {Nikolaev}, {Nordby}, {O'Connor}, {Oguri}, {Oliver}, {Olivier}, {Olsen},
  {Olsen}, {Olszewski}, {Oluseyi}, {Padilla}, {Parker}, {Pepper}, {Peterson},
  {Petry}, {Pinto}, {Pizagno}, {Popescu}, {Prsa}, {Radcka}, {Raddick},
  {Rasmussen}, {Rau}, {Rho}, {Rhoads}, {Richards}, {Ridgway}, {Robertson},
  {Roskar}, {Saha}, {Sarajedini}, {Scannapieco}, {Schalk}, {Schindler},
  {Schmidt}, {Schmidt}, {Schneider}, {Schumacher}, {Scranton}, {Sebag},
  {Seppala}, {Shemmer}, {Simon}, {Sivertz}, {Smith}, {Allyn Smith}, {Smith},
  {Spitz}, {Stanford}, {Stassun}, {Strader}, {Strauss}, {Stubbs}, {Sweeney},
  {Szalay}, {Szkody}, {Takada}, {Thorman}, {Trilling}, {Trimble}, {Tyson}, {Van
  Berg}, {Vand en Berk}, {VanderPlas}, {Verde}, {Vrsnak}, {Walkowicz}, {Wand
  elt}, {Wang}, {Wang}, {Warner}, {Wechsler}, {West}, {Wiecha}, {Williams},
  {Willman}, {Wittman}, {Wolff}, {Wood-Vasey}, {Wozniak}, {Young}, {Zentner},
  \& {Zhan}}]{lsst2009a}
{LSST Science Collaboration}, {Abell}, P.~A., {Allison}, J., {et~al.} 2009,
  arXiv e-prints, arXiv:0912.0201.
\newblock \doarXiv{0912.0201}

\bibitem[{{Lusso} \& {Risaliti}(2017)}]{luss2017a}
{Lusso}, E., \& {Risaliti}, G. 2017, \aap, 602, A79,
  \dodoi{10.1051/0004-6361/201630079}

\bibitem[{{MacLeod} {et~al.}(2015){MacLeod}, {Morgan}, {Mosquera}, {Kochanek},
  {Tewes}, {Courbin}, {Meylan}, {Chen}, {Dai}, \& {Chartas}}]{macl2015a}
{MacLeod}, C.~L., {Morgan}, C.~W., {Mosquera}, A., {et~al.} 2015, \apj, 806,
  258, \dodoi{10.1088/0004-637X/806/2/258}

\bibitem[{{McHardy} {et~al.}(2014){McHardy}, {Cameron}, {Dwelly}, {Connolly},
  {Lira}, {Emmanoulopoulos}, {Gelbord}, {Breedt}, {Arevalo}, \&
  {Uttley}}]{mcha2014a}
{McHardy}, I.~M., {Cameron}, D.~T., {Dwelly}, T., {et~al.} 2014, \mnras, 444,
  1469, \dodoi{10.1093/mnras/stu1636}

\bibitem[{{McLeod} {et~al.}(1998){McLeod}, {Bernstein}, {Rieke}, \&
  {Weedman}}]{mcle1998a}
{McLeod}, B.~A., {Bernstein}, G.~M., {Rieke}, M.~J., \& {Weedman}, D.~W. 1998,
  \aj, 115, 1377, \dodoi{10.1086/300285}

\bibitem[{{Mediavilla} {et~al.}(2011){Mediavilla}, {Mu{\~n}oz}, {Kochanek},
  {Guerras}, {Acosta-Pulido}, {Falco}, {Motta}, {Arribas}, {Manchado}, \&
  {Mosquera}}]{medi2011a}
{Mediavilla}, E., {Mu{\~n}oz}, J.~A., {Kochanek}, C.~S., {et~al.} 2011, \apj,
  730, 16, \dodoi{10.1088/0004-637X/730/1/16}

\bibitem[{{Morgan} {et~al.}(2018){Morgan}, {Hyer}, {Bonvin}, {Mosquera},
  {Cornachione}, {Courbin}, {Kochanek}, \& {Falco}}]{morg2018a}
{Morgan}, C.~W., {Hyer}, G.~E., {Bonvin}, V., {et~al.} 2018, \apj, 869, 106,
  \dodoi{10.3847/1538-4357/aaed3e}

\bibitem[{{Morgan} {et~al.}(2008){Morgan}, {Kochanek}, {Dai}, {Morgan}, \&
  {Falco}}]{morg2008a}
{Morgan}, C.~W., {Kochanek}, C.~S., {Dai}, X., {Morgan}, N.~D., \& {Falco},
  E.~E. 2008, \apj, 689, 755, \dodoi{10.1086/592767}

\bibitem[{{Morgan} {et~al.}(2010){Morgan}, {Kochanek}, {Morgan}, \&
  {Falco}}]{morg2010a}
{Morgan}, C.~W., {Kochanek}, C.~S., {Morgan}, N.~D., \& {Falco}, E.~E. 2010,
  \apj, 712, 1129, \dodoi{10.1088/0004-637X/712/2/1129}

\bibitem[{{Mortonson} {et~al.}(2005){Mortonson}, {Schechter}, \&
  {Wambsganss}}]{mort2005a}
{Mortonson}, M.~J., {Schechter}, P.~L., \& {Wambsganss}, J. 2005, \apj, 628,
  594, \dodoi{10.1086/431195}

\bibitem[{{Mosquera} \& {Kochanek}(2011)}]{mosq2011a}
{Mosquera}, A.~M., \& {Kochanek}, C.~S. 2011, \apj, 738, 96,
  \dodoi{10.1088/0004-637X/738/1/96}

\bibitem[{{Motta} {et~al.}(2012){Motta}, {Mediavilla}, {Falco}, \&
  {Mu{\~n}oz}}]{mott2012a}
{Motta}, V., {Mediavilla}, E., {Falco}, E., \& {Mu{\~n}oz}, J.~A. 2012, \apj,
  755, 82, \dodoi{10.1088/0004-637X/755/1/82}

\bibitem[{{Motta} {et~al.}(2017){Motta}, {Mediavilla}, {Rojas}, {Falco},
  {Jim{\'e}nez-Vicente}, \& {Mu{\~n}oz}}]{mott2017a}
{Motta}, V., {Mediavilla}, E., {Rojas}, K., {et~al.} 2017, \apj, 835, 132,
  \dodoi{10.3847/1538-4357/835/2/132}

\bibitem[{{Mr{\'o}z} {et~al.}(2017){Mr{\'o}z}, {Udalski}, {Skowron}, {Poleski},
  {Koz{\l}owski}, {Szyma{\'n}ski}, {Soszy{\'n}ski}, {Wyrzykowski},
  {Pietrukowicz}, {Ulaczyk}, {Skowron}, \& {Pawlak}}]{mroz2017a}
{Mr{\'o}z}, P., {Udalski}, A., {Skowron}, J., {et~al.} 2017, \nat, 548, 183,
  \dodoi{10.1038/nature23276}

\bibitem[{{Mu{\~n}oz} {et~al.}(2016){Mu{\~n}oz}, {Vives-Arias}, {Mosquera},
  {Jim{\'e}nez-Vicente}, {Kochanek}, \& {Mediavilla}}]{muno2016a}
{Mu{\~n}oz}, J.~A., {Vives-Arias}, H., {Mosquera}, A.~M., {et~al.} 2016, \apj,
  817, 155, \dodoi{10.3847/0004-637X/817/2/155}

\bibitem[{{Navarro} {et~al.}(1997){Navarro}, {Frenk}, \& {White}}]{nava1997a}
{Navarro}, J.~F., {Frenk}, C.~S., \& {White}, S. D.~M. 1997, \apj, 490, 493,
  \dodoi{10.1086/304888}

\bibitem[{{Peterson} {et~al.}(2004){Peterson}, {Ferrarese}, {Gilbert}, {Kaspi},
  {Malkan}, {Maoz}, {Merritt}, {Netzer}, {Onken}, {Pogge}, {Vestergaard}, \&
  {Wandel}}]{pete2004a}
{Peterson}, B.~M., {Ferrarese}, L., {Gilbert}, K.~M., {et~al.} 2004, \apj, 613,
  682, \dodoi{10.1086/423269}

\bibitem[{{Poindexter} \& {Kochanek}(2010{\natexlab{a}})}]{poin2010a}
{Poindexter}, S., \& {Kochanek}, C.~S. 2010{\natexlab{a}}, \apj, 712, 658,
  \dodoi{10.1088/0004-637X/712/1/658}

\bibitem[{{Poindexter} \& {Kochanek}(2010{\natexlab{b}})}]{point2010a}
---. 2010{\natexlab{b}}, \apj, 712, 658, \dodoi{10.1088/0004-637X/712/1/658}

\bibitem[{{Poindexter} {et~al.}(2008){Poindexter}, {Morgan}, \&
  {Kochanek}}]{poin2008a}
{Poindexter}, S., {Morgan}, N., \& {Kochanek}, C.~S. 2008, \apj, 673, 34,
  \dodoi{10.1086/524190}

\bibitem[{{Pooley} {et~al.}(2007){Pooley}, {Blackburne}, {Rappaport}, \&
  {Schechter}}]{pool2007a}
{Pooley}, D., {Blackburne}, J.~A., {Rappaport}, S., \& {Schechter}, P.~L. 2007,
  \apj, 661, 19, \dodoi{10.1086/512115}

\bibitem[{{Pooley} {et~al.}(2019){Pooley}, {Anguita}, {Bhatiani}, {Chartas},
  {Cornachione}, {Dai}, {Fian}, {Mediavilla}, {Morgan}, {Moustakas},
  {Mukherjee}, {O'Dowd}, {Rojas}, {Sluse}, {Vernardos}, \&
  {Webster}}]{pool2019a}
{Pooley}, D., {Anguita}, T., {Bhatiani}, S., {et~al.} 2019, \baas, 51, 411.
\newblock \doarXiv{1904.12968}

\bibitem[{{Refsdal} {et~al.}(2000){Refsdal}, {Stabell}, {Pelt}, \&
  {Schild}}]{refs2000a}
{Refsdal}, S., {Stabell}, R., {Pelt}, J., \& {Schild}, R. 2000, \aap, 360, 10.
\newblock \doarXiv{astro-ph/0005371}

\bibitem[{{Rojas} {et~al.}(2014){Rojas}, {Motta}, {Mediavilla}, {Falco},
  {Jim{\'e}nez-Vicente}, \& {Mu{\~n}oz}}]{roja2014a}
{Rojas}, K., {Motta}, V., {Mediavilla}, E., {et~al.} 2014, \apj, 797, 61,
  \dodoi{10.1088/0004-637X/797/1/61}

\bibitem[{{Rojas} {et~al.}(2020){Rojas}, {Motta}, {Mediavilla},
  {Jim{\'e}nez-Vicente}, {Falco}, \& {Fian}}]{roja2020a}
---. 2020, \apj, 890, 3, \dodoi{10.3847/1538-4357/ab63cb}

\bibitem[{{Schechter} \& {Wambsganss}(2002)}]{sche2002a}
{Schechter}, P.~L., \& {Wambsganss}, J. 2002, \apj, 580, 685,
  \dodoi{10.1086/343856}

\bibitem[{{Shajib} {et~al.}(2019){Shajib}, {Birrer}, {Treu}, {Auger},
  {Agnello}, {Anguita}, {Buckley-Geer}, {Chan}, {Collett}, {Courbin},
  {Fassnacht}, {Frieman}, {Kayo}, {Lemon}, {Lin}, {Marshall}, {McMahon},
  {More}, {Morgan}, {Motta}, {Oguri}, {Ostrovski}, {Rusu}, {Schechter},
  {Shanks}, {Suyu}, {Meylan}, {Abbott}, {Allam}, {Annis}, {Avila}, {Bertin},
  {Brooks}, {Carnero Rosell}, {Carrasco Kind}, {Carretero}, {Cunha}, {da
  Costa}, {De Vicente}, {Desai}, {Doel}, {Flaugher}, {Fosalba},
  {Garc{\'\i}a-Bellido}, {Gerdes}, {Gruen}, {Gruendl}, {Gutierrez}, {Hartley},
  {Hollowood}, {Hoyle}, {James}, {Kuehn}, {Kuropatkin}, {Lahav}, {Lima},
  {Maia}, {March}, {Marshall}, {Melchior}, {Menanteau}, {Miquel}, {Plazas},
  {Sanchez}, {Scarpine}, {Sevilla-Noarbe}, {Smith}, {Soares-Santos},
  {Sobreira}, {Suchyta}, {Swanson}, {Tarle}, \& {Walker}}]{shaj2019a}
{Shajib}, A.~J., {Birrer}, S., {Treu}, T., {et~al.} 2019, \mnras, 483, 5649,
  \dodoi{10.1093/mnras/sty3397}

\bibitem[{{Shakura} \& {Sunyaev}(1973)}]{shak1973a}
{Shakura}, N.~I., \& {Sunyaev}, R.~A. 1973, \aap, 500, 33

\bibitem[{{Shalyapin} {et~al.}(2012){Shalyapin}, {Goicoechea}, \&
  {Gil-Merino}}]{shal2012a}
{Shalyapin}, V.~N., {Goicoechea}, L.~J., \& {Gil-Merino}, R. 2012, \aap, 540,
  A132, \dodoi{10.1051/0004-6361/201118316}

\bibitem[{{Shalyapin} {et~al.}(2008){Shalyapin}, {Goicoechea}, {Koptelova},
  {Ull{\'a}n}, \& {Gil-Merino}}]{shal2008a}
{Shalyapin}, V.~N., {Goicoechea}, L.~J., {Koptelova}, E., {Ull{\'a}n}, A., \&
  {Gil-Merino}, R. 2008, \aap, 492, 401, \dodoi{10.1051/0004-6361:200810447}

\bibitem[{{Shen}(2016)}]{shen2016a}
{Shen}, Y. 2016, \apj, 817, 55, \dodoi{10.3847/0004-637X/817/1/55}

\bibitem[{{Sluse} {et~al.}(2012){Sluse}, {Chantry}, {Magain}, {Courbin}, \&
  {Meylan}}]{slus2012a}
{Sluse}, D., {Chantry}, V., {Magain}, P., {Courbin}, F., \& {Meylan}, G. 2012,
  \aap, 538, A99, \dodoi{10.1051/0004-6361/201015844}

\bibitem[{{Steele} {et~al.}(2004){Steele}, {Smith}, {Rees}, {Baker}, {Bates},
  {Bode}, {Bowman}, {Carter}, {Etherton}, {Ford}, {Fraser}, {Gomboc}, {Lett},
  {Mansfield}, {Marchant}, {Medrano-Cerda}, {Mottram}, {Raback}, {Scott},
  {Tomlinson}, \& {Zamanov}}]{stee2004a}
{Steele}, I.~A., {Smith}, R.~J., {Rees}, P.~C., {et~al.} 2004, in Society of
  Photo-Optical Instrumentation Engineers (SPIE) Conference Series, Vol. 5489,
  \procspie, ed. J.~{Oschmann}, Jacobus~M., 679--692, \dodoi{10.1117/12.551456}

\bibitem[{{Suyu} {et~al.}(2013){Suyu}, {Auger}, {Hilbert}, {Marshall}, {Tewes},
  {Treu}, {Fassnacht}, {Koopmans}, {Sluse}, {Bland ford}, {Courbin}, \&
  {Meylan}}]{suyu2013a}
{Suyu}, S.~H., {Auger}, M.~W., {Hilbert}, S., {et~al.} 2013, \apj, 766, 70,
  \dodoi{10.1088/0004-637X/766/2/70}

\bibitem[{{Tewes} {et~al.}(2013){Tewes}, {Courbin}, \& {Meylan}}]{tewe2013a}
{Tewes}, M., {Courbin}, F., \& {Meylan}, G. 2013, \aap, 553, A120,
  \dodoi{10.1051/0004-6361/201220123}

\bibitem[{{van Dokkum}(2001)}]{vand2001b}
{van Dokkum}, P.~G. 2001, \pasp, 113, 1420, \dodoi{10.1086/323894}

\bibitem[{{Vanden Berk} {et~al.}(2001){Vanden Berk}, {Richards}, {Bauer},
  {Strauss}, {Schneider}, {Heckman}, {York}, {Hall}, {Fan}, {Knapp},
  {Anderson}, {Annis}, {Bahcall}, {Bernardi}, {Briggs}, {Brinkmann}, {Brunner},
  {Burles}, {Carey}, {Castander}, {Connolly}, {Crocker}, {Csabai}, {Doi},
  {Finkbeiner}, {Friedman}, {Frieman}, {Fukugita}, {Gunn}, {Hennessy},
  {Ivezi{\'c}}, {Kent}, {Kunszt}, {Lamb}, {Leger}, {Long}, {Loveday}, {Lupton},
  {Meiksin}, {Merelli}, {Munn}, {Newberg}, {Newcomb}, {Nichol}, {Owen}, {Pier},
  {Pope}, {Rockosi}, {Schlegel}, {Siegmund}, {Smee}, {Snir}, {Stoughton},
  {Stubbs}, {SubbaRao}, {Szalay}, {Szokoly}, {Tremonti}, {Uomoto}, {Waddell},
  {Yanny}, \& {Zheng}}]{vand2001a}
{Vanden Berk}, D.~E., {Richards}, G.~T., {Bauer}, A., {et~al.} 2001, \aj, 122,
  549, \dodoi{10.1086/321167}

\bibitem[{{Walsh} {et~al.}(1979){Walsh}, {Carswell}, \& {Weymann}}]{wals1979a}
{Walsh}, D., {Carswell}, R.~F., \& {Weymann}, R.~J. 1979, \nat, 279, 381,
  \dodoi{10.1038/279381a0}

\bibitem[{{Watson} {et~al.}(2011){Watson}, {Denney}, {Vestergaard}, \&
  {Davis}}]{wats2011a}
{Watson}, D., {Denney}, K.~D., {Vestergaard}, M., \& {Davis}, T.~M. 2011,
  \apjl, 740, L49, \dodoi{10.1088/2041-8205/740/2/L49}

\bibitem[{{Wong} {et~al.}(2020){Wong}, {Suyu}, {Chen}, {Rusu}, {Millon},
  {Sluse}, {Bonvin}, {Fassnacht}, {Taubenberger}, {Auger}, {Birrer}, {Chan},
  {Courbin}, {Hilbert}, {Tihhonova}, {Treu}, {Agnello}, {Ding}, {Jee},
  {Komatsu}, {Shajib}, {Sonnenfeld}, {Bland ford}, {Koopmans}, {Marshall}, \&
  {Meylan}}]{wong2020a}
{Wong}, K.~C., {Suyu}, S.~H., {Chen}, G. C.~F., {et~al.} 2020, \mnras,
  \dodoi{10.1093/mnras/stz3094}

\bibitem[{{Xie} {et~al.}(2016){Xie}, {Shao}, {Shen}, {Liu}, \& {Li}}]{xie2016a}
{Xie}, X., {Shao}, Z., {Shen}, S., {Liu}, H., \& {Li}, L. 2016, \apj, 824, 38,
  \dodoi{10.3847/0004-637X/824/1/38}

\end{thebibliography}

\end{document}